\documentclass[twocolumn,prd,nofootinbib,showpacs,floatfix,preprintnumbers]{revtex4}
\usepackage{epsfig}
\usepackage[dvips]{color}

\def\alt{\raise0.3ex\hbox{$\;<$\kern-0.75em\raise-1.1ex\hbox{$\sim\;$}}}
\def\agt{\raise0.3ex\hbox{$\;>$\kern-0.75em\raise-1.1ex\hbox{$\sim\;$}}}

\definecolor{Black}{named}{Black}
\definecolor{Red}{named}{Red}

\newcommand{\bw}{\begin{widetext}}
\newcommand{\ew}{\end{widetext}}
\def\d{{\rm d}}
\begin{document}

\title{Angular Signatures of Annihilating Dark Matter in the Cosmic Gamma-Ray Background}
\author{A.~Cuoco$^{1}$, J.~Brandbyge$^{1}$,  S.~Hannestad$^{1}$, T.~Haugb{\o}lle$^{1,2}$,
G.~Miele$^{3,4}$}
\affiliation{$^1$Department of Physics and Astronomy, University
of Aarhus, Ny Munkegade, Bygn. 1520 8000 Aarhus Denmark
\\$^2$Instituto de F\'{\i}sica Te\'{o}rica UAM-CSIC, Universidad Aut\'{o}noma
de Madrid, Cantoblanco, 28049 Madrid, Spain
\\$^3$Universit\`{a} ``Federico II'', Dipartimento di Scienze
Fisiche, Napoli, Italy \& INFN Sezione di Napoli
\\$^4$Instituto de F\'{\i}sica Corpuscular (CSIC-Universitat de
Val\`{e}ncia), Ed. Institutos de Investigaci\'{o}n, Apdo. 22085,
E-46071 Valencia, Spain}

\date{\today}

\begin{abstract}
The extragalactic cosmic gamma-ray background (CGB) is an
interesting channel to look for signatures of dark matter
annihilation. In particular, besides the imprint in the energy
spectrum, peculiar anisotropy patterns are expected compared to
the case of a pure astrophysical origin of the CGB. We take into
account the uncertainties in the dark matter clustering properties
on sub-galactic scales, deriving two possible anisotropy
scenarios. A clear dark matter angular signature is achieved when
the annihilation signal receives only a \emph{moderate}
contribution from sub-galactic clumps and/or cuspy haloes.
Experimentally, if galactic foregrounds systematics are
efficiently kept under control, the angular differences are
detectable with the forthcoming GLAST observatory, provided that
the annihilation signal contributes to the CGB for a fraction
$\agt$10-20\%. If, instead,  sub-galactic structures have a more
prominent role, the astrophysical and dark matter anisotropies
become degenerate, correspondingly diluting the DM signature. As
complementary observables we also introduce the cross-correlation
between surveys of galaxies and the CGB and the cross-correlation
between different energy bands of the CGB and we find that they
provide a further sensitive tool to detect the dark matter angular
signatures.
\end{abstract}
\preprint{IFT-UAM/CSIC-07-52}
\pacs{95.35.+d, 95.85.Pw, 98.70.Vc}
\maketitle

%%%%%%%%%%%%%%%%%%%%%%%%%%%%%%%%%%%%%%%%%%%%
\section{Introduction}
%%%%%%%%%%%%%%%%%%%%%%%%%%%%%%%%%%%%%%%%%%%%

Astronomical and cosmological observations provide overwhelming
evidence for the presence of dark matter (DM) (see e.g.\
\cite{Bertone:2004pz} for a review). In particular, the combination
of various cosmological data sets provides a precise measurement of
the amount of DM in the universe: $\Omega_c h^2 \simeq 0.11$ with a
$2 \sigma$ precision of $\sim$ 5\% in the minimal $\Lambda$CDM model
\cite{Tegmark:2006az,Spergel:2006hy,Seljak:2006bg} and $\sim$ 20\%
in more extended models \cite{Hamann:2006pf}.

However, despite the noticeable sensitivity to the cosmological
abundance of matter (either dark or baryonic), such measurements
only weakly constrain the properties and nature of the particle
associated to DM, and very weak limits are available on the DM
particle mass $m_\chi$ and on its couplings. The simplest DM
candidate is the Weakly Interacting Massive Particle (WIMP) which
is characterized by having been in thermal equilibrium in the
early universe (as opposed to for example the sterile neutrino or
super-heavy DM), and having decoupled from equilibrium while
non-relativistic. In order to get the correct DM abundance the
mass of such a particle cannot be larger than $\sim 30$ TeV
\cite{Bertone:2004pz,Griest:1989wd}. On the other hand, collider
experiments provide a lower bound on the mass of $\sim 50-100$ GeV
\cite{Bertone:2004pz}, depending on the specific particle
candidate. Mass of $\mathcal{O}$(GeV) are however possible if more
exotic candidates are considered \cite{Gunion:2005rw}. The typical
thermally averaged DM annihilation cross section in the WIMP
scenario is $ <\!\!\! \sigma_\chi v \!\!\! >$$\sim 10 ^{-26}$
cm$^{3}$s$^{-1}$ \cite{Bertone:2004pz}. However, we stress that if
the DM is produced out of equilibrium in the early universe, no
bounds can be given and super-massive, GUT scale, DM particles
($m_\chi \sim 10^{15}$ GeV) and cross sections $ <\!\!\!
\sigma_\chi v \!\!\! >$ $ \ll 10^{-26}$ cm$^{3}$s$^{-1}$ are in
principle possible.

From the point of view of particle physics WIMP candidates are very
appealing and emerge naturally in Supersymmetric (SUSY) extensions
of the standard model or in the Universal Extra Dimensions (UED)
model \cite{Bertone:2004pz}. The sensitivity of accelerator
experiments, notably the Large Hadron Collider, and of direct search
experiments are approaching the levels required to test the WIMP
hypothesis, and a direct discovery of DM WIMPs could happen in the
not so distant future.

DM WIMP candidates have thus typically a large annihilation cross
section and  pair-annihilate into standard model particles that
subsequently decay and shower producing large numbers of photons
and neutrinos. Such $\gamma$-rays  from DM annihilation constitute
an ideal target for astronomical searches. Thus, astrophysical and
cosmological observations can provide a crucial test,
complementary to a direct laboratory detection, in the search for
the nature of DM particles. Various astrophysical environments
have been discussed in detail as promising sites for observation
of DM annihilation, among others the galactic center, satellite
dwarf galaxies of the Milky Way and clumps of DM in the Milky Way
halo. In the following we will focus instead on the all-sky
diffuse signal expected in the extragalactic cosmic gamma-ray
background (CGB)
\cite{Bergstrom:2001jj,Ullio:2002pj,Elsaesser:2004ck,Elsaesser:2004ap}.

Peculiar spectral and angular features can
help in disentangling a signal produced by DM from emission by
``ordinary'' astrophysical sources. The spectrum of photons from
DM annihilation is in general harder than the spectra arising from
normal astrophysical processes and exhibit a pronounced cutoff at
an energy near  $m_\chi$ \cite{Bergstrom:2001jj,Ullio:2002pj}. The
resulting emission thus appear like a ``bump'' in the background
astrophysical energy spectrum in the energy range in which the DM
signal gives a relevant contribution. However, although this kind
of signature would constitute a strong hint of DM annihilation,
astrophysical processes that could mimic such behavior are
possible.

Another signature, which has been widely studied, is direct
annihilation into a state containing photons, resulting in a line in
the background spectrum that would constitute a ``smoking gun''
signature of DM. However, by construction this process is
necessarily loop suppressed and in most models the flux is quite
small (see, however, \cite{Ullio:2002pj} for a more thorough
discussion).

Peculiar angular signatures thus offer a complementary signature
to exclude the remaining degenerate astrophysical interpretations
of a signal.  An example is the clumpiness of DM at sub-galactic
scales
\cite{Berezinsky:2005py,Taylor:2002zd,Pieri:2007ir,Bergstrom:1998jj}
investigated by recent zoomed high-resolution N-body
simulations \cite{Diemand:2006ik,Diemand:2005vz}: Clumpiness would
result in a population of high galactic latitude extended gamma
emitters with a typical annihilating DM gamma spectrum. These
kinds of objects could hardly be associated to astrophysical
emitters (but see \cite{Baltz:2006sv}). In these models the size of the
clumps is expected to have a characteristic distribution and thus the
anisotropy of the integrated signal from all the clumps also
exhibits a characteristic behavior \cite{Pieri:2007ir}.

Likewise, the expected angular anisotropies both in the case of an
astrophysical and of a DM origin of the CGB can be calculated, and
have received increasing attention in the last few years
\cite{Cuoco:2006tr,Ando:2005xg,Ando:2006mt,Ando:2006cr,Miniati:2007ke}.
In the following we will further pursue this issue addressing the
differences expected in the two cases and their detectability in
the light of the improved statistics that will be available, when
the GLAST observatory is launched and start to take data in the
near future. We will compare throughout the paper our findings in
particular with \cite{Ando:2005xg,Ando:2006cr} that deal
specifically with anisotropies induced by DM annihilation.
Already, there have been claims
\cite{Elsaesser:2004ap,deBoer:2005tm} of a DM signal in the CGB as
observed by EGRET (see also
\cite{Bergstrom:2006tk,deBoer:2006ck,deBoer:2007zc}), although
with the limited EGRET statistics and with the uncertainties in
the galactic foregrounds, alternative astrophysical explanations
cannot be ruled out. On the other hand, with the improved
statistics from GLAST, a proper analysis of the anisotropy
properties of the CGB should be able to prove, or disprove, the DM
interpretation of features in the CGB spectrum.

Complementary to previous studies we shall employ in the following
a parametric approach characterizing the expected CGB signal in
terms of a few key parameters, that catch the relevant physical
aspects of the problem, and varying them in order to asses the
robustness and/or model dependence of the possible signatures. A
further advantage of this approach is to make explicit the various
assumptions employed throughout on which the final signatures
depend. The relevant parameters in the following will be the
degree of correlation of the CGB sources with matter and the
absolute normalization of the signal, or, equivalently, the
expected collected statistics. Further, we will also consider
complementary anisotropy observables like the  cross-correlation
between surveys of galaxies and the CGB and  the cross-correlation
between different energy bands of the CGB. Together with the
auto-correlations of the CGB these represent a set of independent
observables that can be jointly employed improving considerably
the sensitivity to the DM signal.

The paper is organized as follows: In section \ref{CGBmain} we
present a discussion of the horizons within which the CGB signal
is expected to come, relevant for the determination of the
intensity of the CGB anisotropies itself. In section
\ref{CGBanisotropies} we introduce the formalism to derive the CGB
anisotropies in terms of the angular power spectrum. In section
\ref{GLASTforecast} we present a forecast for the expected
statistics from GLAST and we discuss the possibility of
disentangling the DM annhilation signal from that of astrophysical
processes. In sections \ref{crosscorr} and \ref{crosscorr_eb} we
introduce the cross-correlation between the CGB and galaxy surveys
and the cross-correlation between different energy bands of the
CGB and similarly we discuss the different behavior and
sensitivity in the two cases of interest. In section
\ref{discussion} we discuss  how the previous conclusions apply to
different possible scenarios for the CGB and DM properties. In
section \ref{final} we summarize and conclude.

%%%%%%%%%%%%%%%%%%%%%%%%%%%%%%%%%%%%%%%%%%%%
\section{Gamma-Ray Horizons}\label{CGBmain}

%%%%%%%%%%%%%%%%%%%%%%%%%%%%%%%%%%%%%%%%%%%%
%\subsection{Gamma-Ray Horizons}
%%%%%%%%%%%%%%%%%%%%%%%%%%%%%%%%%%%%%%%%%%%%

The extragalactic cosmic gamma-ray signal can be parameterized as
\cite{Ullio:2002pj,Cuoco:2006tr}
\begin{equation}
I(E_\gamma,\hat{n})\propto \int_0^\infty \!\!\!\d z\,
\frac{\rho^\alpha(z,\hat{n},r(z))\,g[E_\gamma(1+z)]\,e^{-\tau(E_\gamma,z)}}{H(z)\,(1+z)^{3}}
\,,\label{intcosmo}
\end{equation}
where  $g(E)=dN_\gamma/dE$ is the photon spectrum of the sources,
$E_\gamma$ is the energy we observe today, $\rho(z,\hat{n},r)$ is
the matter density in the direction $\hat{n}$ at a comoving
distance $r$, and the redshift $z$ is used as time variable. In
the following we will interchangeably  use $\rho$, or $\rho_\chi$
when we want to underline the particle nature of DM. The sources
are assumed to be distributed proportional to $\rho^\alpha$. The
Hubble expansion rate is related to its present $z\!=\!0$ value
$H_0$ through the matter and the cosmological constant energy
densities as $H(z)= H_0\sqrt{\Omega_M(1+z)^3+\Omega_\Lambda}$, and
the reduced Hubble expansion rate $h(z)$ is given by $H(z)=100 \
h(z)$ km/s/Mpc. We will in the following use the parameters of the
standard $\Lambda$CDM model \cite{Spergel:2006hy}, i.e. $\Omega_M
= 0.3$, $\Omega_\Lambda = 0.7$ and $H_0 = 70$ km/s/Mpc. The
quantity $\tau(E_\gamma,z)$ is the optical depth of photons to
absorptions via pair production (PP) on the  Extra-galactic
Background Light (EBL). In ref.~\cite{Cuoco:2006tr} an energy
threshold of $E_{\rm cut}=100$ GeV has been considered resulting
in a PP horizon of about $z\approx 0.5$, and a simple
extrapolation back in time of the present EBL gave a sufficiently
accurate value of $\tau$. In the present work we also consider
$E_{\rm cut}=10$ GeV and horizons as large as $z\approx 4-5$. In
this range the dynamical evolution of the EBL during the photon
propagation becomes important for a correct estimate of $\tau$. To
take this into account we use the parametrization of
$\tau(E_\gamma,z)$  from \cite{Stecker:2005qs} for
$0\!<\!z\!<\!5$, where evolution effects are included in the
calculation. The EBL is expected to be negligible at redshifts
higher than $z\approx 5$ corresponding to the peak of star
formation. Thus, gamma photons produced at earlier times
experience an undisturbed propagation until $z\approx 5$, while
only in the recent epoch they start to loose energy, due to
scattering on the EBL. Correspondingly, we assume
$\tau(E_\gamma,z)=\tau(E_\gamma,5)$ for $z>5$ (see also formula
(A.6) in \cite{Cuoco:2006tr}).

The case $\alpha=1$ is generally representative of astrophysical
sources following the Large Scale Structure (LSS) of matter, while
the case $\alpha=2$ is appropriate for annihilating DM whose
signal follows the square of the matter density, $\propto
\rho_\chi^2$ through
\begin{equation}
I_{\chi} = \frac{ <\!\!\!  \sigma_\chi v \!\!\!
>}{8\pi m^2_\chi}  \int_0^\infty \!\!\!\d z\,
\frac{\rho_\chi^2(z,\hat{n},r(z))\,g[E_\gamma(1+z)]\,e^{-\tau(E_\gamma,z)}}{H(z)\,(1+z)^{3}}
\,.\label{intcosmoDM}
\end{equation}
This last point is however entangled with the exact small scale
(sub-galactic) clustering properties of DM and deserves further
discussions. If DM clumps on sub-galactic scales, as suggested by
various numerical models of galaxy formation, or if the DM halo
has a very pronounced spike at its center, the galactic DM signal
can be greatly enhanced and the overall cosmological contribution
of DM to the CGB would be due to the emission from single
galaxies. The $\rho_\chi^2(\vec{x})$ field in Eq.(2) would look
approximately as a sum of delta functions centered on the
galaxies' positions and the DM annihilation signal would thus
trace the matter distribution linearly, (actually, the galaxy
distribution), at least at scales larger than the galactic haloes.
In this case, however, the DM signal expected from the Milky Way
itself would probably be a more promising observable for
signatures of DM annihilation, as we will further discuss later.
The relative contribution of the galactic versus extra-galactic DM
signal is further discussed in \cite{Hooper:2007be}. In principle,
if the DM clustering properties would be known in the whole range
from sub-galactic to cosmological scales, the ratio of the linear
to quadratic contribution can be calculated. However, given the
still persisting uncertainty in the sub-galactic clumping, to be
general we will assume a DM annihilation anisotropy signal $\delta
I_\chi/I_\chi \propto \rho_\chi^2 / \overline{\rho_\chi^2} +\xi \
\rho_\chi / \bar{\rho}_\chi$, where $\xi$ parameterizes the
relative weights of the linear and quadratic contributions. In the
following we will discuss mainly the extreme scenarios $\xi\ll1$
and $\xi\gg1$ in which one of the two contributions dominates over
the other. More precisely, we thus define a ``quadratic scenario''
in which DM clustering  is relevant only above the scale of
galactic haloes ($\sim 10^{12} M_\odot$), and a ``linear
scenario'' in which sub-galactic structures dominate the
cosmological DM annihilation signal. The mixed scenario is
discussed further in section \ref{discussion}.

\begin{figure}[t]
\vspace{-1.0pc}
\begin{center}
\begin{tabular}{c}
\vspace{-1.5pc}\includegraphics[width=0.50\textwidth]{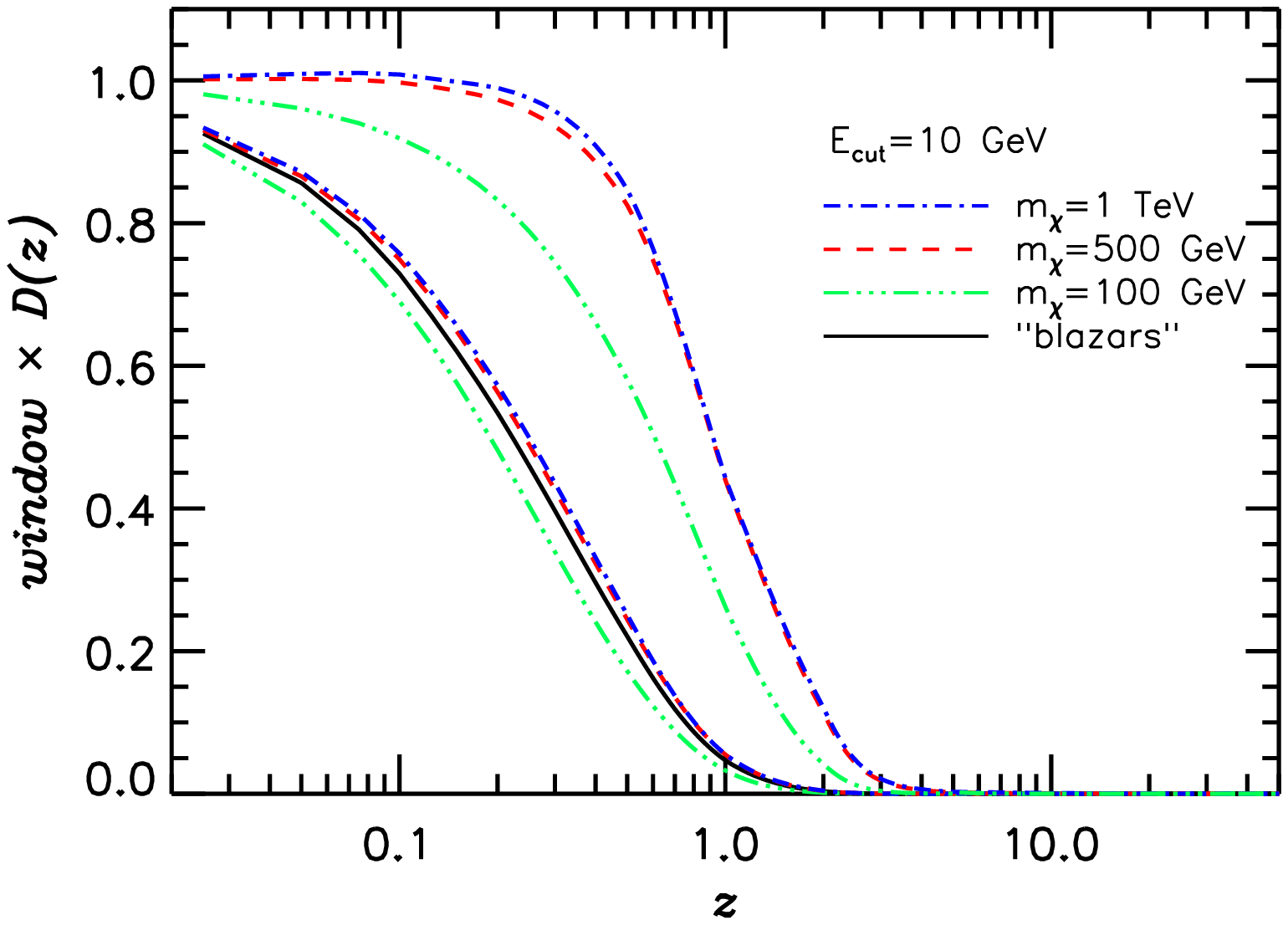} \\
\vspace{-1.5pc}\includegraphics[width=0.50\textwidth]{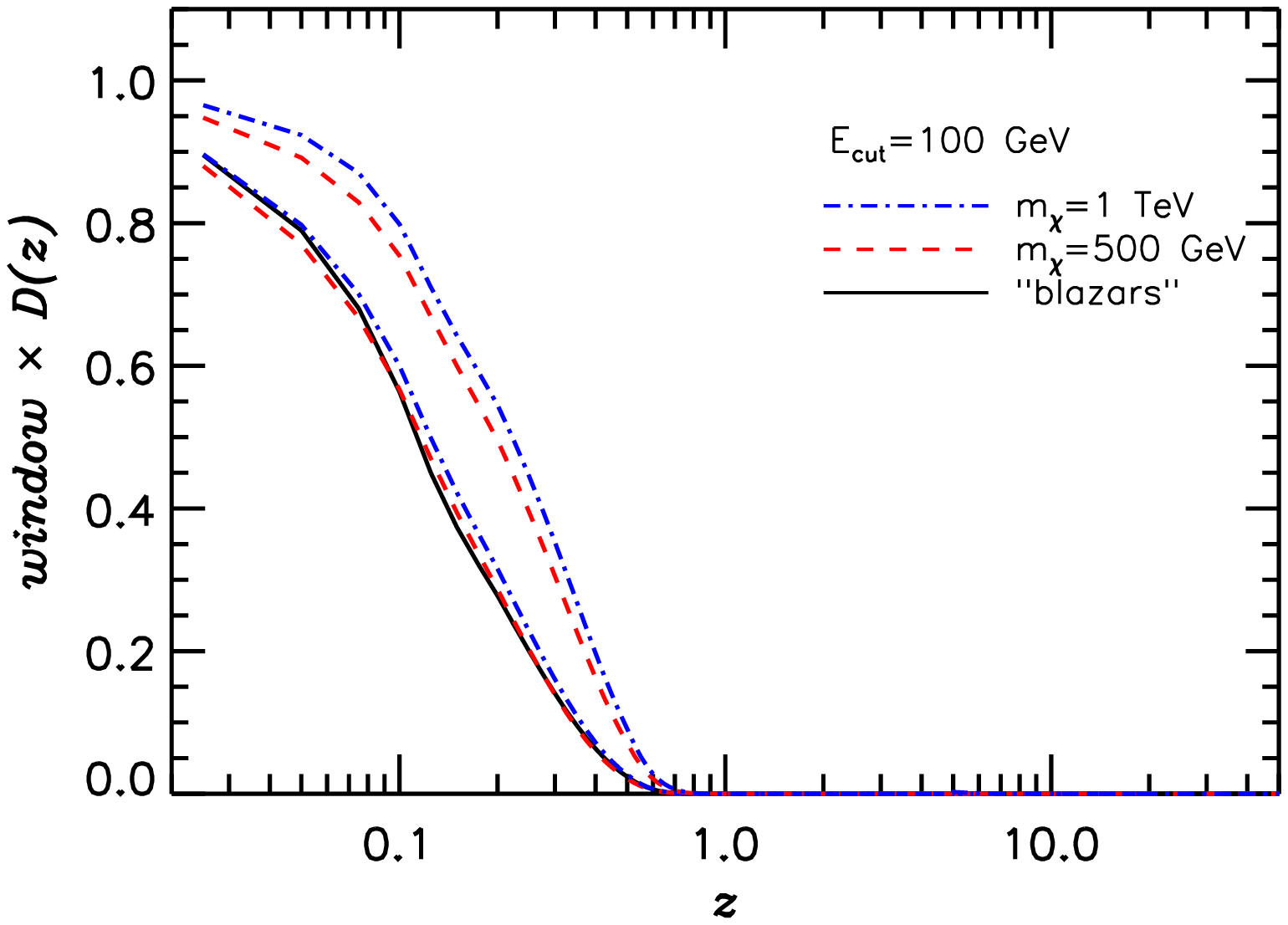}
\end{tabular}
\vspace{-1.5pc}
\end{center}
\hspace{-1.5pc} \caption{Gamma window functions times the linear
growth factor $D(z)$ for $E_{\rm cut}$ = 10, 100 GeV and for
various DM masses. The curves are normalized to 1 at $z=0$. Upper
and lower curves in each panel refer to the DM annihilation signal
correlating quadratically and linearly ($\alpha=2,1$),
respectively, with matter. \label{fig:windows}}
\end{figure}

The astrophysical and DM window functions, $W_{\gamma}(E_{\gamma
\rm cut},z)$ and $W_{\chi}(E_{\chi \rm cut},z)$, which contain the
information about gamma-ray propagation, injection spectra and
cosmological effects, are defined from Eq.~(\ref{intcosmo}) as
\begin{eqnarray}\label{gammaflux1}
\!\!\!\!\!I_{\gamma}(E_{\gamma \rm cut},\hat{n}) &\propto&\!
\int_0^\infty \!\!\!\d z\, W_{\gamma}(E_{\gamma \rm cut},z) \,
\rho (z,\hat{n})\,, \\
\label{gammaflux2} \!\!\!\!\!I_{\chi}(E_{\chi \rm cut},\hat{n})
&\propto&\! \int_0^\infty \!\!\!\d z\, W_{\chi}(E_{\chi \rm
cut},z) \, \rho^2 (z,\hat{n})\,,
\end{eqnarray}
where we are using the notation $\rho (z,\hat{n},r(z))=(1\!+\!z)^3
\times \rho (z,\hat{n})$ to underline that the window function is
only dependent on the two variables, direction and redshift, and
to make explicit the $(1\!+\!z)^3$ behavior of the matter density.
In principle $\rho_s$, the density distribution of astrophysical
sources, should be used in Eq.~(\ref{gammaflux1}): $\rho_s$ in
general exhibits  a scale and time dependent bias with respect to
the matter density. However, specific classes of astrophysical
gamma-ray sources have different  biases. Blazars, for example,
that most likely produce the bulk of the CGB signal detected by
EGRET, are well known to concentrate at the center of clusters of
galaxies, thus presenting an over-bias with respect to galaxies at
high densities. On the other hand, galaxies and clusters of
galaxies quite fairly trace the matter density, at least in the
recent cosmic epoch. The assumption $\rho_s=\rho$ for $I_{\gamma}$
is thus general enough to reasonably describe emission from
astrophysical sources.

The window functions can be found from Eq.~(\ref{intcosmo}) and are given by
\begin{equation}
W(E_{\rm cut},z)\!\equiv\!\int_{E_{\rm cut}}^{\infty}\!\!\!\!\!\!\!\d E\,
 \frac{ g[E(1+z)] \,(1+z)^{3\alpha-3}\!\!}{H(z)} \, e^{-\tau(E,z)}\!,
\label{gammawindow}
\end{equation}
where $\alpha=1,2$ applies in the astrophysical and DM cases,
respectively. When properly normalized  $W(E_{\rm cut},z)$
represents the probability of receiving a photon of $E_\gamma>
E_{\rm cut}$ emitted at a redshift $z$. It can be used to define
an effective horizon, $z_{\cal H}$, beyond which the probability
of receiving a photon is negligible (e.g.$\alt 1\%$). For $E_{\rm
cut}\agt 100$ GeV PP losses dominate and the horizon is $z_{\cal
H}\alt 1$ independent of the value of $\alpha$ or the shape of
$g(E)$. For $E_{\rm cut}\alt 10$ GeV, instead, PP losses start to
become negligible ($\tau\approx 0$) and photons propagate freely
from arbitrary high redshifts. However, even in this case a
horizon exists due to redshifting related this time to the exact
shape of the injection spectrum $g(E)$ and the value of $\alpha$.
In the case of astrophysical sources we take $g(E)\propto E^{-2}$,
consistent with the observed EGRET CGB spectrum and with the
observed spectra of common astrophysical gamma sources like
blazars. We found however that for $E_{\rm cut}=10$ GeV  the
horizon is still mainly settled by the cosmological and PP
attenuation effects while the exact shape of the spectrum plays a
minor role and even choices like $g(E)\propto E^{-1}$ or
$g(E)\propto E^{-3}$ change only slightly the astrophysical
window. Given the poor sensitivity to the specific details of the
emission spectrum we will thus often refer in the following to the
term ``\emph{blazars}'', meaning in general a representative class
of astrophysical gamma-emitters tracing linearly the matter
density and with a power law  $E^{-2}$ spectrum. The resulting
horizon is $z_{\cal H}\approx 1$ as shown in
Fig.~\ref{fig:windows}. The windows are further multiplied by the
linear growth factor $D(z)$ that takes into account the evolution
of matter clustering in the past (see the next section). $D(z)$
gives a further, although not crucial, contribution to the
determination of the exact horizon $z_{\cal H}$. For $E_{\rm
cut}=100$ GeV the horizon is instead $z_{\cal H}\approx 0.5$ and
depends exclusively on the EBL absorption both in the
astrophysical and DM cases. This makes this energy range
particularly interesting due to its limited sensitivity to any
particular modelling. Some further effects can in fact contribute
to modify the horizon: The luminosity of blazars for example can
in principle change with time due to well known source evolution
effects introducing a further $(1\!+\!z)^\lambda$ factor in the
window. While evolution effects are unimportant for $E_{\rm
cut}=100$ GeV, a strong source evolution can in principle affect
$z_{\cal H}$  at $E_{\rm cut}=10$ GeV.

In the case of DM, the spectrum $g(E)$ and the cosmological factors
involved are quite different and the effective horizon can
be much larger. The different expected horizon is in fact
an important ingredient for a clear discrimination through the
expected pattern and intensity of the anisotropies. A commonly used
parametrization for the annihilation spectrum of DM is
\cite{Bergstrom:2001jj}
\begin{equation}\label{DMspectrum}
  g(E)\propto \frac{ \exp(-7.76
  E/M_{\chi})}{(E/M_{\chi})^{1.5}+0.00014},
\end{equation}
i.e.~a spectrum that is generally harder than the astrophysical
$E^{-2}$ spectrum, and with a cutoff near the DM mass energy (that
is the behavior responsible for the bump in the overall spectrum).
The shape given by Eq.~(\ref{DMspectrum}) is almost independent of
the details of the annihilation process, at least for the case of
SUSY WIMPs where the main contribution comes from decays to
$q\bar{q}$, $ZZ$, and  $W^+W^-$, with subsequent hadronization. A
slightly different spectrum is expected for the case of decay into
a lepton-anti-lepton pair or for the annihilation of UED WIMPs
(see \cite{Bertone:2004pz} for details). We will not further
consider these cases although basically our findings also apply to
them. The resulting windows depend on the assumed DM mass,
$m_{\chi}$, and on the chosen $E_{\rm cut}$. Various cases used in
the following are shown in Fig.~\ref{fig:windows}. At energies
above $E \agt$ 100 GeV the photon absorption dominates and, as
discussed above, the DM and astrophysical horizons are almost
identical, $z_{\cal H}\approx 0.5$. For $E_{\rm cut}$ = 10 GeV the
horizon for astrophysical sources  is $z_{\cal H} \simeq 1$, while
that of DM is generally of order $z_{\cal H} \simeq 3-4$. Very
interestingly, we can see that the role of absorption by the EBL
is still quite relevant for DMA at $E_{\rm cut}$ = 10 GeV limiting
the horizon which otherwise would exceed $z \simeq 10$ giving much
smaller DM anisotropies. Finally, for the case of the DM signal
correlating linearly with matter ($\alpha=1$), no appreciable
differences are present in the windows neither at the high nor at
the low energy cut. Even if the DM spectral shape is quite
peculiar, not unexpectedly this seems to play a minor role, as in
the previously discussed case of astrophysical emission. In this
case the DM and astrophysical signal have degenerate anisotropy
properties and this observable cannot further help in
disentangling the two contributions.

%%%%%%%%%%%%%%%%%%%%%%%%%%%%%%%%%%%%%%%%%%%%
\section{CGB Anisotropies}\label{CGBanisotropies}
%%%%%%%%%%%%%%%%%%%%%%%%%%%%%%%%%%%%%%%%%%%%

%%%%%%%%%%%%%%%%%%%%%%%%%%%%%%%%%%%%%%%%%%%%
\subsection{3D Power Spectra}
%%%%%%%%%%%%%%%%%%%%%%%%%%%%%%%%%%%%%%%%%%%%

\begin{figure}[t]
\begin{center}
\vspace{-1pc}\hspace{-1pc}\includegraphics[width=0.50\textwidth]{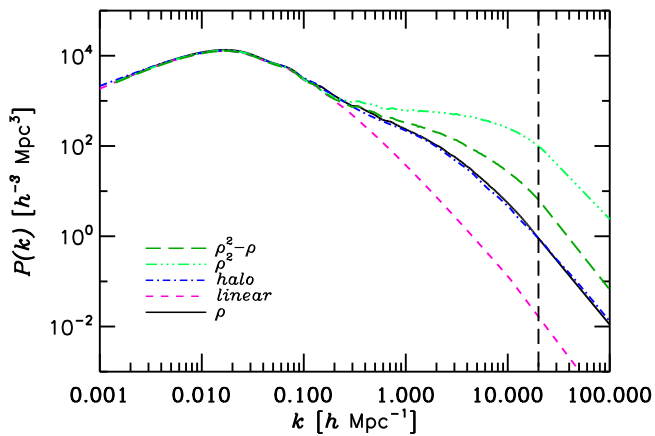}
\end{center}
\vspace{-1pc} \caption{3D power spectra of the matter density
distribution and of its square, as derived from a cosmological
N-body simulation. Also shown is the linear matter power spectrum,
the non-linear Halo-model prediction and the cross-correlation
between the matter distribution and its square. The vertical
dashed line mark the confidence limit on the calculation of
$P(k)$.  All the spectra are normalized at the linear scales to
the matter power spectrum. \label{fig:3Dspectra}}
\end{figure}

To derive the CGB anisotropies we need first to know the spatial
clustering properties of the matter field $\rho$ and of its
square. To this purpose we use a template of the matter
distribution derived from a DM N-body simulation.

The N-body simulation was performed with the publicly available
code \textsc{gadget}-2 \cite{Springel:2005mi} with $512^3$ CDM
particles in a $128\, {\rm Mpc}/h$ box. We have assumed  a flat
$\Lambda$CDM-model, with  $\Omega_{\rm CDM}=0.30$,
$\Omega_\Lambda=0.70$ and $h=0.70$ as well as a scale-invariant
primordial power spectrum, $P\propto{k}$. The transfer function
was generated using CMBFAST \cite{cmbfast}, and then the initial
conditions were computed using second order Lagrangian
perturbation theory \cite{Crocce:2006ve}. The smoothed density
field is constructed by interpolating the particles to a $2048^3$
grid, enforcing mass conservation, and using the adaptive spline
kernel from \cite{monaghan}.

If $\rho(\vec{x})$ denotes the simulation density field and
$\rho(\vec{k})$ its Fourier transform, then the matter power
spectrum can be written as
\begin{equation}
    P_{{\rho}}(k)= \int_{\! \mathcal{S}_{\! \Delta k}} \!\!\!\! d^3\!\vec{k} \, \left| \rho^*(\vec{k}) \rho(\vec{k})
    \right|,
\end{equation}
and analogously
\begin{equation}
    P_{\rho^2}(k)= \int_{\! \mathcal{S}_{\! \Delta k}} \!\!\!\! d^3\!\vec{k} \, \left|  \rho_2^*(\vec{k}) \rho_2(\vec{k}))
    \right|,
\end{equation}
where $\rho_{2}(\vec{k})$ denotes the Fourier transform of the
\emph{squared} density field $\rho^2(\vec{x})$ and $ \mathcal{S}_{\!
\Delta k}$ is a spherical shell of radius $k$ and thickness ${\Delta
k}$. Finally, it is also possible to estimate the cross-correlation
spectrum
\begin{equation}
     P_{\rho\rho^2}(k)= \int_{\! \mathcal{S}_{\! \Delta k}} \!\!\!\! d^3\!\vec{k} \, \left| \rho_2^*(\vec{k})  \rho(\vec{k})
     \right|.
\end{equation}
We take into account the time dependence of $P_{i}(k,z)$
($i=\rho,\rho^2,\rho\rho^2$) using the linear growth factor $D(z)$
\begin{equation}
    P_{i}(k,z)= P_{i}(k,z=0)\times D^2(z),
\end{equation}
with $D(z)\propto h(z)\int_z^{\infty} dz'(1\!+\!z')/h^3(z')$ and
$D(0)=1$.\footnote{$D(z)$ is a good approximation also at
non-linear scales where $P(k,z)$ grows only slightly faster than
the linear growth \cite{Jenkins:1997en}.}

In Fig.~\ref{fig:3Dspectra} the various spectra are shown. Notice
the increase in power at small scales for $P_{\rho^2}(k)$ compared
to $P_{\rho}(k)$. For reference $P_{\rho}(k)$ as calculated in the
Halo-model \cite{Smith:2002dz} is also shown. It can be seen that
the spectra from the N-body simulation and the Halo-model are in
quite good agreement. However, the N-body spectrum starts to be
affected by numerical noise beyond $k\simeq20 \, h$ Mpc$^{-1}$,
shown as a vertical line in the plot, and this range is accordingly
excluded from the analysis. The contribution from higher wave
numbers, $k\agt 20 \, h$ Mpc$^{-1}$, or, equivalently, smaller
scales, $\lambda \lesssim 2 \pi /20 \, h^{-1}$ Mpc, is in any case
relevant only for very high multipoles $l\agt 1000$ not accessible
experimentally, so that for the present purposes they can be safely
neglected.  The spectrum of the squared matter distribution is also
in fair agreement with the Halo-model calculation as derived in
\cite{Ando:2005xg}. The most noticeable feature is an increase in
the intensity of the anisotropies at the non-linear scales $k\agt
1.0 \, h$ Mpc$^{-1}$ with respect to the matter spectrum,
understandable in the framework of the Halo-model as a dominant
contribution from the single-halo term. As expected the
cross-correlation is in between the matter and the matter squared
spectra. In the figure all the spectra are normalized to the matter
spectrum at linear scales, while the absolute normalization for the
matter squared and for the cross-correlation is given by 4 and 2
times this value, respectively.

%%%%%%%%%%%%%%%%%%%%%%%%%%%%%%%%%%%%%%%%%%%%
\subsection{Angular anisotropies}
%%%%%%%%%%%%%%%%%%%%%%%%%%%%%%%%%%%%%%%%%%%%

From Eq.(\ref{gammaflux1})-(\ref{gammaflux2}) we can now easily
construct the angular power spectra of the various dimensionless
fluctuation fields $\delta I/I$
\begin{eqnarray}
  C^l_{\gamma} & = &  \int \frac{dr}{r^2} \; W_{\gamma}^2(r) \;
    P_{\rho}\left(k=\frac{l}{r},z(r)\right)\, , \\
  C^l_{{\chi1}} & = &  \int \frac{dr}{r^2} \; W_{{\chi1}}^2(r) \;
    P_{\rho}\left(k=\frac{l}{r},z(r)\right)\, , \\
  C^l_{{\chi2}} & = &  \int \frac{dr}{r^2} \; W_{{\chi2}}^2(r) \;
    P_{\rho^2}\left(k=\frac{l}{r},z(r)\right)\, ,
\end{eqnarray}
for the astrophysical and DM cases following linearly or
quadratically the matter distribution, respectively.\footnote{The
angular spectra calculations involve an integral over $r$, the
comoving distance, while the windows are known in terms of the
redshift $z$. We thus use the $r$-$z$ relation $r(z)=c/H_0
\int_0^z dz' 1/\sqrt{\Omega_m(1+z')^3+\Omega_{\Lambda}}$, and its
inverse $z(r)$. } We have used the Limber approximation, which is
accurate for all but the very lowest multipoles.

In principle the intermediate case of a DM signal $\propto
\rho_\chi^2 / \overline{\rho_\chi^2} +\xi \ \rho_\chi /
\bar{\rho}_\chi$ can also easily be derived, giving a final spectrum
\begin{equation}
   C^l_{{\chi}}  =  \frac{1}{\left(1+\xi\right)^2} C^l_{{\chi1}} +
                        \frac{\xi^2}{\left(1+\xi\right)^2} C^l_{{\chi2}} +
                         \frac{2\xi}{\left(1+\xi\right)^2}
                         C^l_{{\chi
                        12}} \, ,
\end{equation}
where the $\rho$-$\rho^2$ spectrum is involved
\begin{equation}
      C^l_{{\chi12}}  =   \int \frac{dr}{r^2} \; W_{{\chi1}}(r) W_{{\chi2}}(r) \;
    P_{\rho\rho^2}\left(k=\frac{l}{r},z(r)\right)\, , \\
\end{equation}
and $0\leq\xi\leq\infty$ weights the relative contribution of the
linear and quadratic correlation  terms. In practice, however, in
the following we will mainly consider the two cases
$\xi=0,\infty$, while the intermediate case is easily
understandable with a qualitative discussion. We will however
consider quantitatively a mixed scenario in section
\ref{discussion}.

\begin{figure}[t]
\begin{center}
\vspace{-1pc}
\includegraphics[width=0.50\textwidth]{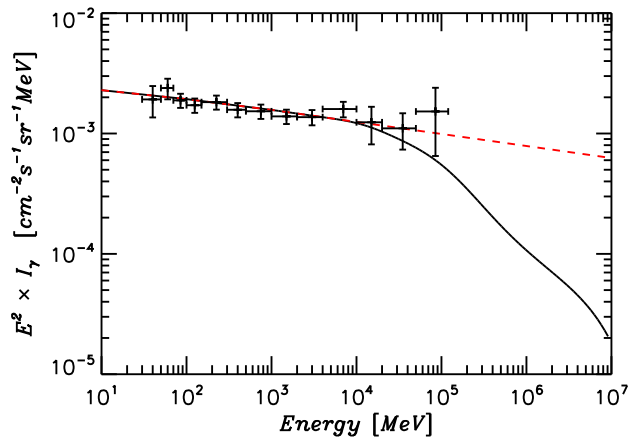}
\end{center}
\vspace{-1pc} \caption{EGRET spectrum from \cite{Sreekumar:1997un}
and extrapolation up to 10 TeV. The solid line shows the expected
effect of the PP attenuation. \label{fig:egrb}}
\end{figure}

\section{CGB auto-correlation analysis}\label{GLASTforecast}

\subsection{Forecast assumptions and sensitivity}

In this section we describe our assumptions to assess the
sensitivity of the forthcoming gamma-ray detectors, in particular
GLAST, to the angular signatures in the auto-correlation spectrum
described in the previous section. Similarly, the sensitivity of the
cross-correlation observables is discussed in the next sections.

The observed diffuse gamma emission is constituted by the sum of
the CGB and of the diffuse galactic emission, so that,
experimentally, the relevant extragalactic signal needs to be
separated from the related galactic foregrounds. We will assume in
the following a perfect removal of the galactic gammma foregrounds
from the CGB. We will thus quote statistical errors only. Indeed,
foreground separation will be a non trivial issue in the analysis
of the forthcoming datasets and  this is turn  is expected to
propagate to the determination of the CGB anisotropies. A detailed
analysis of the effects of the foregrounds is beyond the scope of
this work. We will however further discuss this point in section
\ref{discussion}.

We consider the diffuse energy spectrum as measured by EGRET
\cite{Sreekumar:1997un}
\begin{equation}
I(E_\gamma)=k_0 \left(\frac{E}{0.451 {\rm GeV}}\right)
^{-2.10\pm0.03}\, , \label{spectrum98}
\end{equation}
valid from $E\sim\,$10 MeV to $E\sim\,$100 GeV, where
$k_0=(7.32\pm 0.34)\times 10^{-6} {\rm cm}^{-2}{\rm s}^{-1}{\rm
sr}^{-1}{\rm GeV}^{-1}$, correcting it by the effects of EBL
absorption as described in section \ref{CGBmain}. We show in
Fig.~\ref{fig:egrb} a plot of the EGRET data
\cite{Sreekumar:1997un} together with the fit in
Eq.~(\ref{spectrum98}) and its extrapolation in the case of PP
absorption. In agreement with the result of  section \ref{CGBmain}
it can be seen again that $\sim$10 GeV is the critical energy
above which the EBL absorption effects become relevant.

It is then possible to estimate the number of events, $N_\gamma$,
in the relevant energy range to be collected during the time $t$
as
\begin{equation}
N_\gamma = t\cdot  DC\cdot \Omega_{\rm fov}\cdot \int_{E_{\rm
cut1}}^{E_{\rm cut2}} {\rm d}E\, A_{\rm eff}(E) I_\gamma(E) \,,
\label{Ngamma}
\end{equation}
where $DC$ is the duty-cycle of the instrument, $\Omega_{\rm fov}$
is the solid angle of the field of view and $A_{\rm eff}(E)$ is the
effective collecting area of the instrument (averaged over the field
of view of the instrument). For GLAST \cite{GLAST} we will assume a
constant $A_{\rm eff}(E)=10^4 \, \rm cm^2$, $DC=90\%$, $\Omega_{\rm
fov}=2.4 \, \rm sr$ and $f_{\rm fov}= \Omega_{\rm fov}/4\pi$. In
addition, we use the angular resolution of the experiment
($\sigma_b=0.115^\circ$) and the associated angular window function
$W_l=\exp \left( - l^2 \sigma_b^2 /2\right)$.

Analogously, the number of photons expected from DM annihilation is
given by
\begin{equation}
N_{\chi} = t\cdot  DC\cdot \Omega_{\rm fov}\cdot \int_{E_{\rm
cut1}}^{E_{\rm cut2}} {\rm d}E\, A_{\rm eff}(E) I_{\chi}(E) \,,
\label{NDM}
\end{equation}
where $I_{\chi}(E)$ is the  DM annihilation spectrum. In our
parametric approach we calculate the statistics $N_{\chi}$
normalizing the DM spectrum to the EGRET spectrum so that
$N_{\gamma}=N_{\chi}$ for the relevant energy cut. Later we will
discuss briefly how the conclusions are affected in  the case in
which the DM signal is reduced to the 20\% of the EGRET value or
in the case in which the CGB itself is reduced if parts of it are
resolved as sources by GLAST.

The forecasted error bars on the various  CGB angular
auto-correlation spectra are given by
%
%\bw
\begin{eqnarray} \label{gausserrors1}
  \frac{\delta C^l_{\gamma}}{C^l_{\gamma}} &=&
  \sqrt{\frac{2(1+{C_{N,\gamma}}/W_l^2 C^l_{\gamma})^2}{(2l+1)\Delta l f_{\rm fov}}}, \\
\label{gausserrors2}
  \frac{\delta C^l_{\chi}}{C^l_{\chi}} &=&
  \sqrt{\frac{2(1+{C_{N,\chi}}/W_l^2 C^l_{\chi})^2}{(2l+1)\Delta l f_{\rm
  fov}}},
\end{eqnarray}
%\ew
where $C_{N, \gamma}=\Omega_{\rm fov}/N_{\gamma}$ and $C_{N,
\chi}=\Omega_{\rm fov}/N_{\chi}$ are the gamma and DM random noise
levels respectively.

\begin{figure*}[!t]
\begin{center}
\begin{tabular}{cc}
\includegraphics[width=0.50\textwidth]{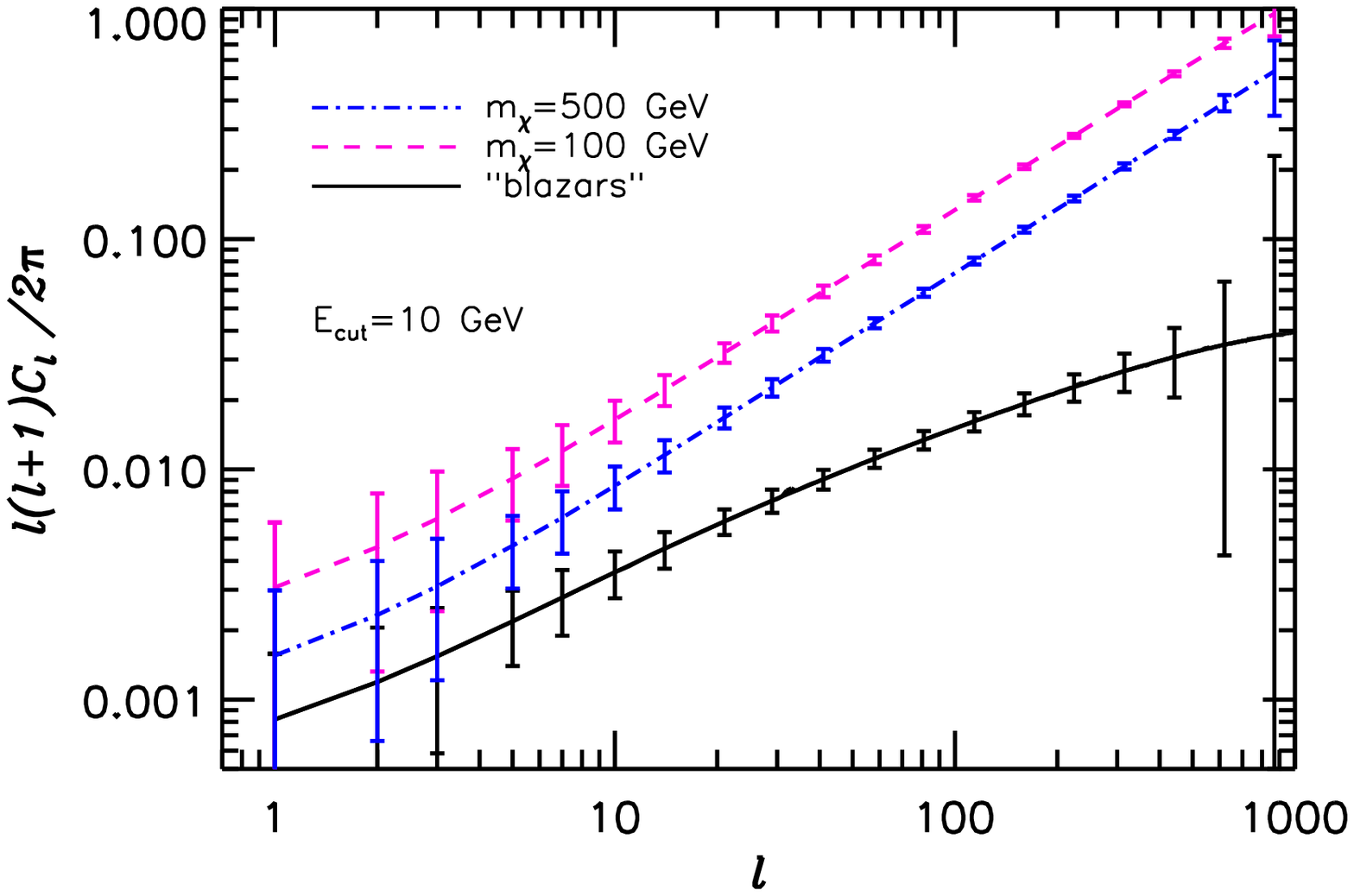}  &
\includegraphics[width=0.50\textwidth]{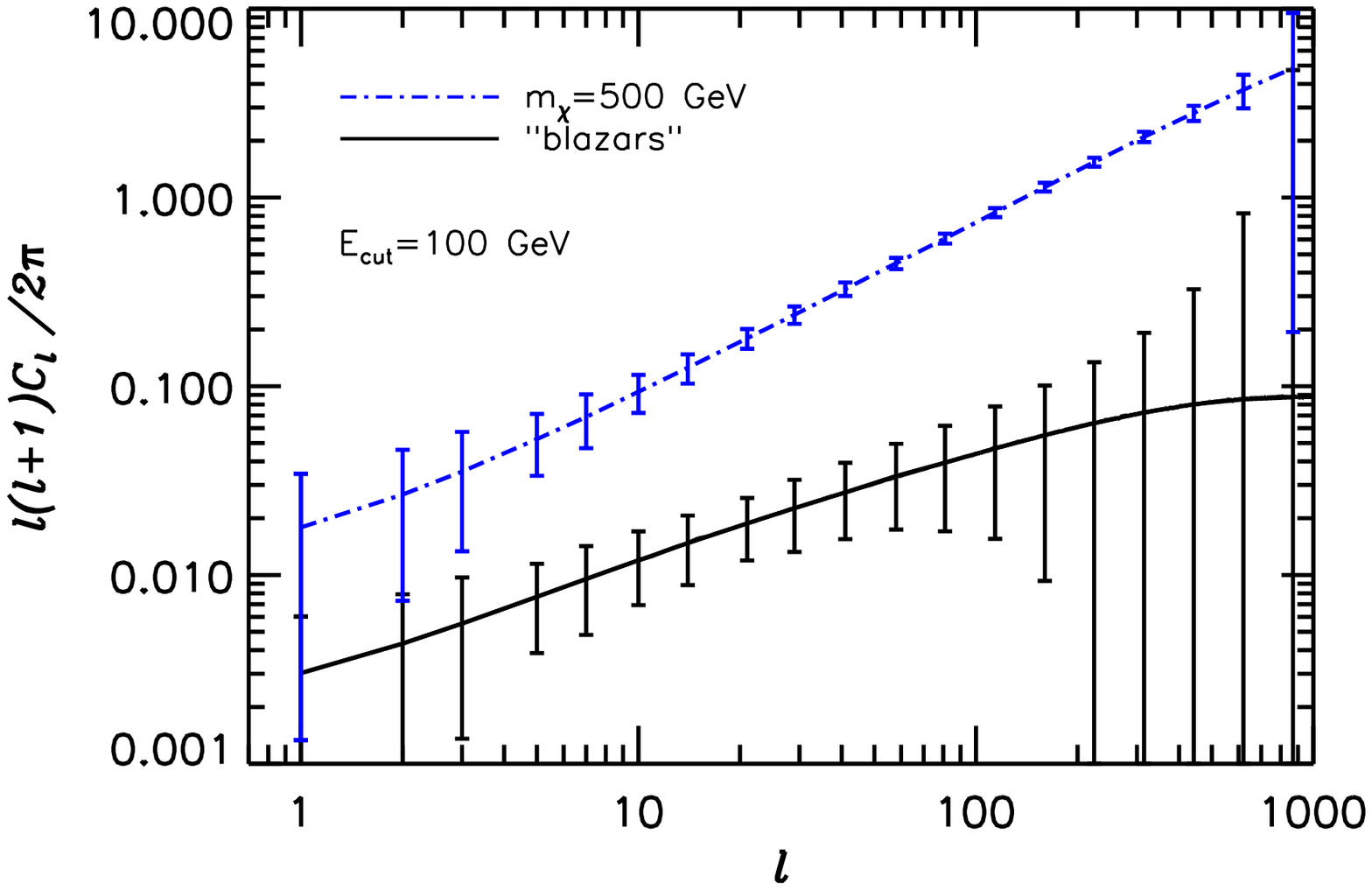}
\end{tabular}
\end{center}
\vspace{-1.5pc} \caption{Angular spectra  for $E_{\rm cut}=10,100$
GeV with $1\ \sigma$ error bars for a 4-year GLAST survey.
\label{fig:angularspectra2} }
\end{figure*}

The resulting spectra and their error bars are shown in
Fig.~\ref{fig:angularspectra2} for the case of the pure quadratic
scenario. In the case of $E_{\rm cut}$ = 10 GeV DM masses of 100,
500 GeV are shown, while for $E_{\rm cut}$ = 100 GeV we consider
only the value 500 GeV, a mass of 1 TeV showing basically the same
spectrum and error bars.

The plot on the right ($E_{\rm cut} = 100$ GeV) shows
quantitatively what was anticipated in the previous section. The
windows are almost identical for the DM and the astrophysical
cases and the higher intrinsic level of fluctuations of DM
produces a much higher normalization in the angular power
spectrum. For reasonable values of the DM mass the change in the
level of anisotropies is thus measurable and distinguishable from
the astrophysical case and provides an important signature of DM
emission.  Further, the shapes of the angular spectra are quite
different, the DM case giving a further enhancement of the
fluctuations at small scales $l\agt 100$ as previously found also
in \cite{Cuoco:2006tr} (see in particular Fig. 2). The statistics
collected above 100 GeV by GLAST in a 4 year period
($N_{\gamma}\approx 10^4$) is still high enough to allow a more
than satisfactory measurement and separation of the various power
spectra.

In the $E_{\rm cut} = 10$ GeV case there is instead a competition
between the enhanced level of fluctuations and their dilution in
the wider horizon related to DM. The final normalization of the
$C_l$'s is still greater though, than that of blazars although the
difference is reduced with respect to $E_{\rm cut} = 100$ GeV. The
increased statistics at low energy and, more importantly, the
different shapes of the spectra, however, still make the two
contributions separable. For $E_{\rm cut} = 100$ GeV, relevant in
the case of a not too light DM particle $m_\chi \agt 300$ GeV, we
further see that the angular spectrum maintains its diagnostic
power with the additional advantage that the small horizon
involved, $z_{\cal H}\approx 0.5$, considerably reduces the model
dependence of the signature from cosmological evolution or bias
effects.

Finally, an important point to consider is that the astrophysical
sources' power spectrum, being almost independent of the energy
cut, can be measured at low energies, where the collected
statistics is high and thus the statistical errors are
correspondingly small. This calibration of the astrophysical
signal at low energies can further improve the separation of the
two signals especially in the case where the DM flux is not at the
EGRET level but significantly below the CGB flux. The amount of
separation can be quantified by considering the
\emph{cross-correlation} between different energy bins. We will
further discuss this point in section \ref{crosscorr_eb}.

\begin{figure*}[t]
\begin{center}
\begin{tabular}{cc}
\includegraphics[width=0.50\textwidth]{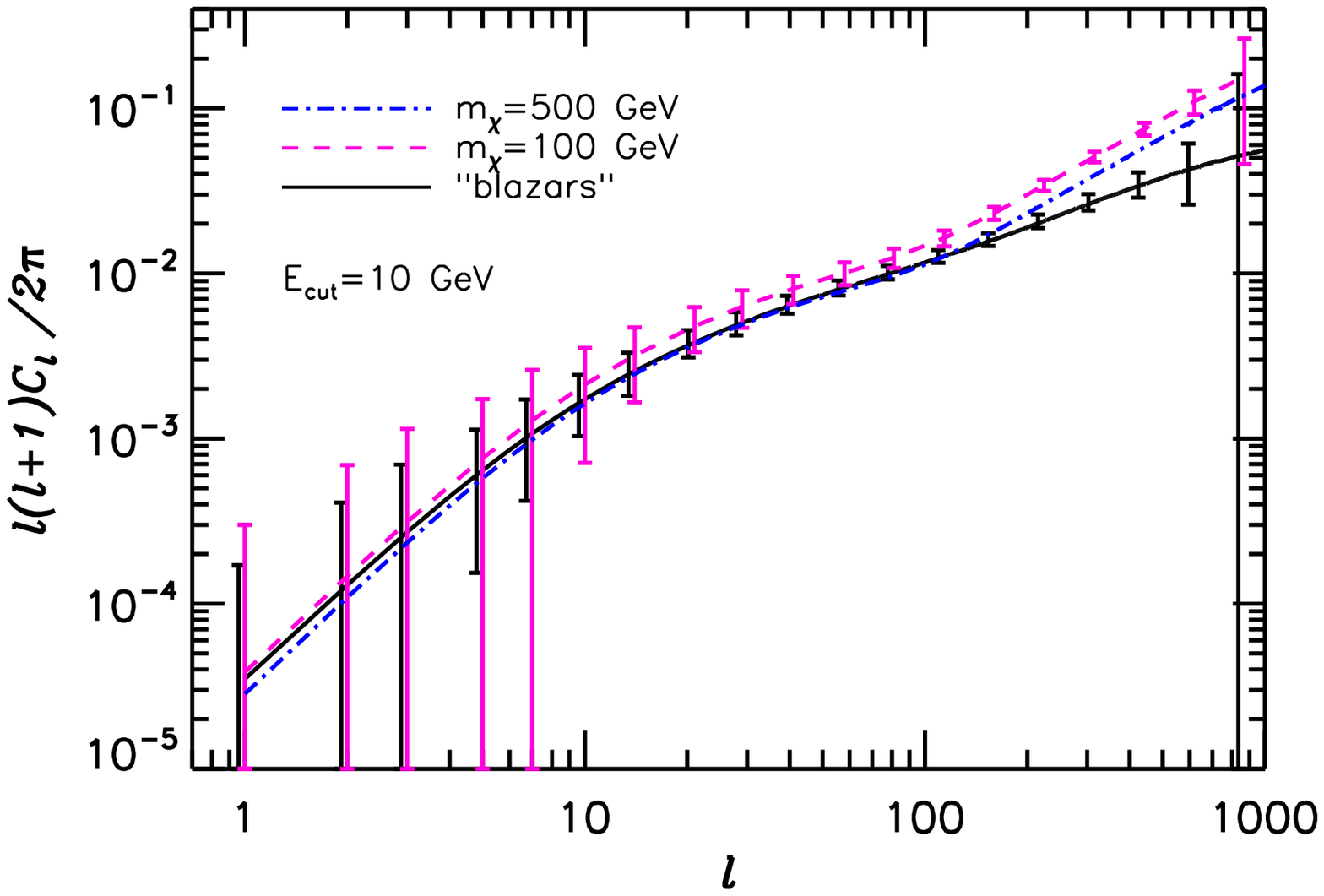}  &
\includegraphics[width=0.50\textwidth]{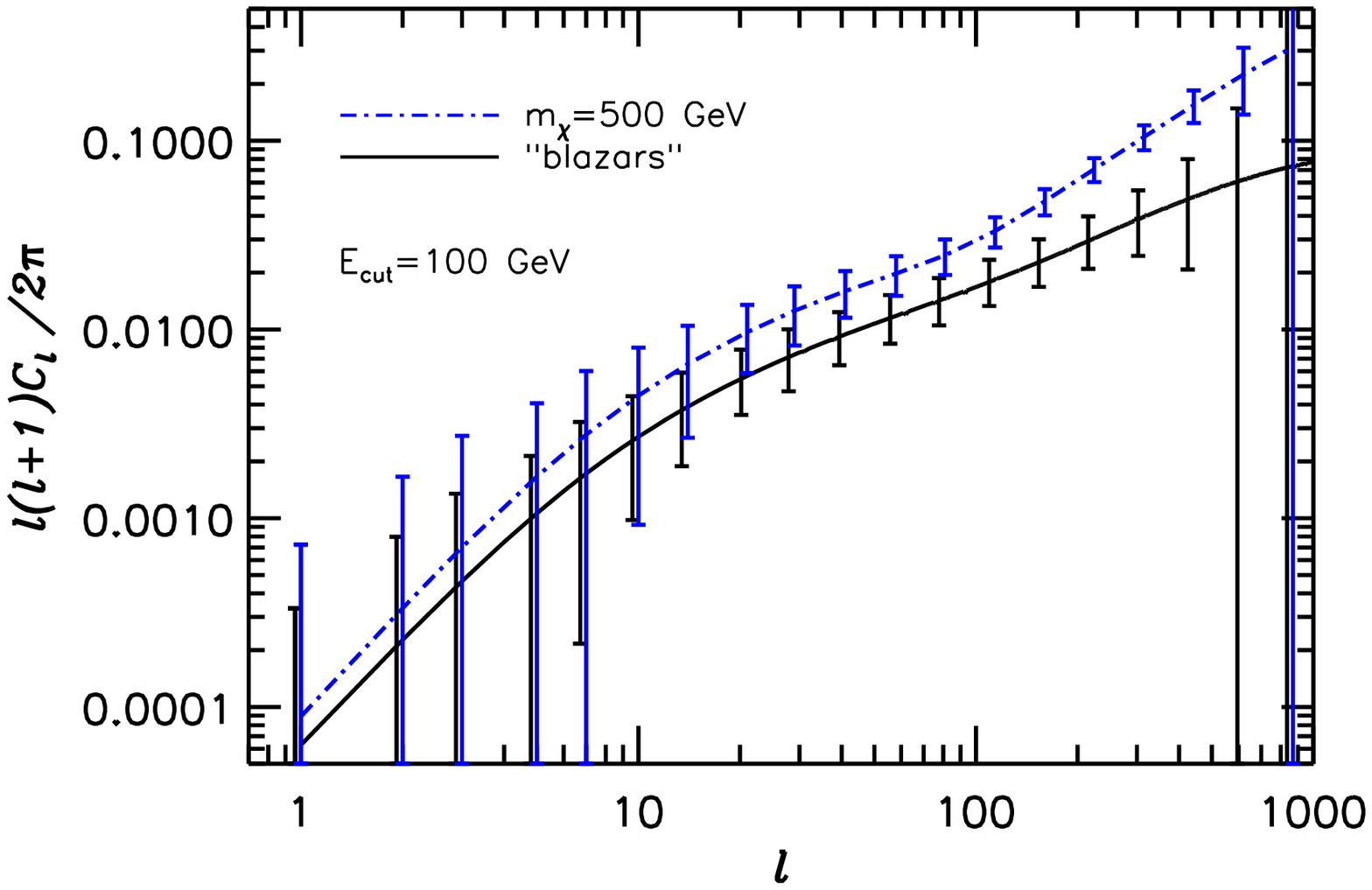}
\end{tabular}
\end{center}
\vspace{-1.5pc} \caption{Angular cross-correlation spectra between
the CGB and an example survey of galaxies for $E_{\rm cut}$ = 10
and 100 GeV. Error bars are for a 4-year GLAST survey.
\label{fig:angularspectra1} }
\end{figure*}

\subsection{Comparison with previous works}

The above results for $E_{\rm cut} = 10$ GeV are in general in
good agreement with \cite{Ando:2005xg,Ando:2006cr}, confirming the
sensitivity of the auto-correlation spectrum to the DM signal. In
particular we confirm that apart the normalization, the blazar and
DM spectra have a quite different shape with the DM case giving
much more power to the small scales, i.e. for multipoles $l\agt
100$.

In the present work, with respect to \cite{Ando:2005xg} we
consider in much more detail the role of photon absorption showing
that it is quite relevant also for an energy cut as low as 10 GeV.
We have indeed also improved the treatment of the photon
absorption process considering the most updated results from
\cite{Stecker:2005qs}.

To compare directly our results with that of \cite{Ando:2005xg} it
should be taken into account that our quadratic model shown in
Fig.~\ref{fig:angularspectra2} consider the contribution to
anisotropies from haloes greater than  average galactic haloes,
with a typical mass of $10^{12} M_\odot$. Ref. \cite{Ando:2005xg},
instead, consider two particular fiducial models with DM
clustering until a sub-halo mass scale of $10^{6} M_\odot$ and
$10^{-6} M_\odot$. Both their model thus consider a certain
sub-galactic contribution and can be approximately compared to our
mixed scenario (see section \ref{discussion}) with a particular
value of the mixing parameter $\xi$. As expected, indeed, the
anisotropy spectra in \cite{Ando:2005xg} has a lower normalization
corresponding to the fact the DM ``linear'' contribution tends to
drag the fluctuations to the level of the astrophysical ones. A
further part of the difference could also arise from our improved
treatment of the photon absorption although in this case, as
explained above, a direct comparison is difficult.

%%%%%%%%%%%%%%%%%%%%%%%%%%%%%%%%%%%%%%%%%%%%
\section{Galaxy-CGB cross-correlation}\label{crosscorr}
%%%%%%%%%%%%%%%%%%%%%%%%%%%%%%%%%%%%%%%%%%%%

Another observable sensitive to the DM properties can be obtained
by looking at the cross-correlation between the CGB and galaxy
catalogues. If the CGB is cosmological in origin, clearly a
positive cross-correlation is expected. Comparing the
cross-correlation originating from DM annihilation to that of
astrophysical emission differences are expected, similar to those
of the auto-correlation spectrum studied in the previous sections.
The same formalism can be generalized to address these differences
in detail as we show in the following. Intuitively, the use of the
cross-correlation spectrum is a way to go beyond the level of the
statistical information only and the limits imposed by cosmic
variance exploiting not only the statistical spectrum $C_l$ but
also the information contained in the whole sky distribution in
terms of the $a_{lm}$ harmonic coefficients \cite{Cuoco:2006tr}.

Similarly to the CGB we introduce the galaxy intensity map of the
catalogue
\begin{equation} I_{g}(\hat{n})= \int_0^\infty \d
z\,  \frac{dn}{dz}(z)   \, \rho_{g}(z,\hat{n}),
\end{equation}
where the galaxy window $W_{g}(z)=dn/dz(z)$ is in this case the
redshift distribution of the catalogue's galaxies. The related
observables in this case are the cross-correlation between gamma
emission and galaxies and the DM-galaxy cross-correlation
\begin{eqnarray}
\!\!\!\!C^l_{\gamma g} & = &  \int \frac{dr}{r^2} \;
W_{\gamma}(r)W_g(r) \;
    P_{\rho}\left(k\!=\!\frac{l}{r},z(r)\right)\, , \\
\!\!\!\!C^l_{{\chi1} g} & = &  \int \frac{dr}{r^2} \;
W_{{\chi1}}(r)W_g(r) \;
    P_{\rho}\left(k\!=\!\frac{l}{r},z(r)\right)\, ,  \\
\!\!\!\!C^l_{{\chi2} g} & = &  \int \frac{dr}{r^2} \;
W_{{\chi2}}(r)W_g(r) \;
    P_{\rho\rho^2}\left(k\!=\!\frac{l}{r},z(r)\right)\, .
\end{eqnarray}
As a simplifying hypothesis we again neglect the matter-galaxy
bias. Notice further that in a galaxy catalogue galaxies are
observed directly so that the $\rho_g$ from the catalogue already
contains the redshift evolution and no further $(1\!+\!z)^3$
factors are needed. The function $W_{g}(z)=dn/dz(z)$ is
characteristic of the survey and of its depth i.e.~the mean
observed redshift. In the following we will assume the typical
shape
\begin{equation}
  W_{g}(z)=\frac{dn}{dz}(z)=   \left( z-z_c  \right)^2    \exp \left[ -\left( \frac{z-z_c}{z_0-z_c}
  \right)^{1.5} \right] \, ,
\end{equation}
where $z_0$ is the mean redshift depth of the survey and $z_c$ is
the low $z$ cutoff. For definiteness we will consider
a 2MASS-like catalogue with $z_0=0.1$ and $z_c=0$.

The error bars for these observables are this time given by a more
involved expression \bw
\begin{eqnarray}
  \frac{\delta C^l_{\gamma g}}{C^l_{\gamma g}} &=&
  \sqrt{\frac{1}{(2l+1)\Delta l f_{\rm fov}}
  \left( 1+      \frac{ C^l_{\gamma}\; C^l_{g}}{C^l_{\gamma g} \; C^l_{\gamma g}}
  \left(1+{C_{N,\gamma}}/W_l^2 C^l_{\gamma}) (1+{C_{N,g}}/W_l^2 C^l_{g}
  \right)
     \right)} \ ,  \\
  \frac{\delta C^l_{\chi g}}{C^l_{\chi g}} &=&
  \sqrt{\frac{1}{(2l+1)\Delta l f_{\rm fov}}
  \left( 1+      \frac{ C^l_{\chi}\; C^l_{g}}{C^l_{\chi g} \; C^l_{\chi g}}
  \left(1+{C_{N,\chi}}/W_l^2 C^l_{\chi}) (1+{C_{N,g}}/W_l^2 C^l_{g}
  \right)
     \right)} \ ,
\end{eqnarray}
\ew where $C_{N,g}=\Omega_{\rm fov}/N _{g}$ is the galaxy random
noise, analogous to  $C_{N, \gamma}$ and  $C_{N, \chi}$, where
$N_g$ is the number of galaxies in the survey. For the case of the
2MASS survey we assume $f_{sky}\simeq 0.8$ and $N_g\simeq 10^6$.
We have further assumed that the CGB and galaxy maps have been
smoothed to the same angular resolution so that the same $W_l$ can
be used. The use of cross-correlation with galaxies has been
proven to be a powerful tool in cosmology, in particular in the
analysis of the Cosmic Microwave Background Radiation
\cite{Afshordi:2003xu,Cabre:2006qm}. A cross-correlation with
galaxies has also been suggested in the study of the MeV gamma
background \cite{Zhang:2004tj}. We refer the reader to these
references for further details on the formalism employed.

The results are shown in Fig.~\ref{fig:angularspectra1} with the
same assumptions as in Fig.~\ref{fig:angularspectra2} for
$N_{\gamma}$ and $N_{\chi}$. It can be seen that in general the
galaxy-CGB spectrum is less optimal with respect to the
auto-correlation spectrum of the CGB itself to look for
differences between DM and astrophysical sources, but there are
still some discerning power. The same trend as in
Fig.~\ref{fig:angularspectra2} is recognizable: At $E_{\rm
cut}=100$ GeV the fluctuations in the DM spectrum are higher with
respect to the astrophysical case and the statistics and angular
resolution expected from GLAST can distinguish the two cases. At
$E_{\rm cut}=100$ GeV the balance between enhanced DM fluctuations
and horizon dilution makes degenerate the normalization of the two
contributions. However, the different shapes at $l\agt 100$ still
allow to disentangle the two cases. Notice that in all cases the
intermediate scale multipoles $l\sim100$ appear to be optimal to
disentangle the two cases.

Although in the single case shown the cross-correlation appears to
be less sensitive to DM signatures compared to the auto-correlation
of the CGB, it has to be stressed that this is an independent
observable and the two can be combined and used at the same time
improving the statistical power of the analysis. Further, different
catalogues can be employed with, possibly, a more suitable window
that can enhance the sensitivity of the cross-correlation. Finally,
if the catalogue is sufficiently deep (like for example the case of
the SDSS main sample and the high redshift Luminous Red Galaxies
sample \cite{Cabre:2006qm}) it can be possible to split the galaxy
distribution into various redshift bins and perform a tomography
analysis with different independent cross-correlations.

%%%%%%%%%%%%%%%%%%%%%%%%%%%%%%%%%%%%%%%%%%%%
\section{Cross-correlation between energy bands}\label{crosscorr_eb}
%%%%%%%%%%%%%%%%%%%%%%%%%%%%%%%%%%%%%%%%%%%%

\begin{figure}[t]
\vspace{-1pc}
\begin{center}
\vspace{0pc}
\begin{tabular}{c}
\includegraphics[width=0.50\textwidth]{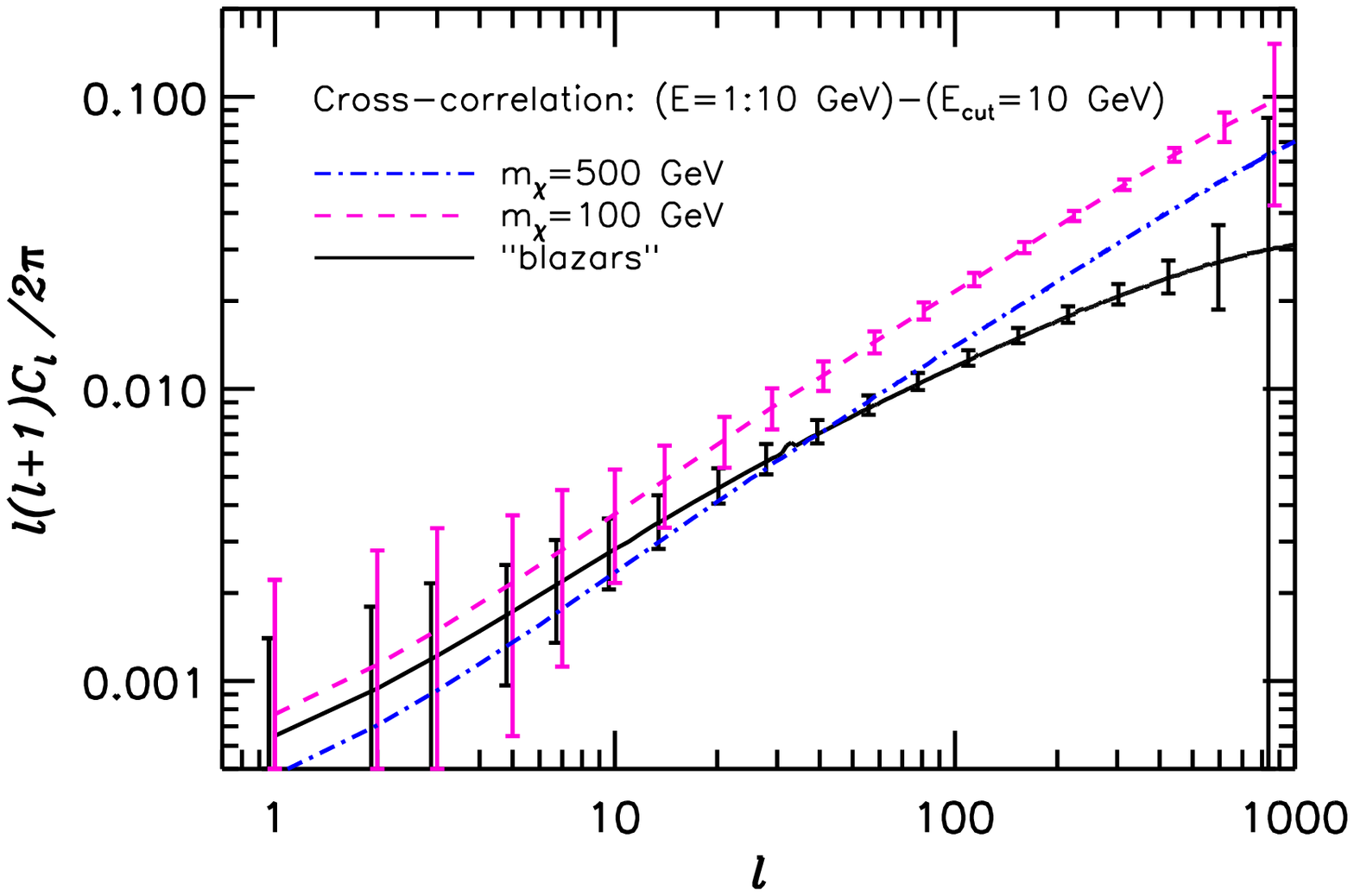} \\
\includegraphics[width=0.50\textwidth]{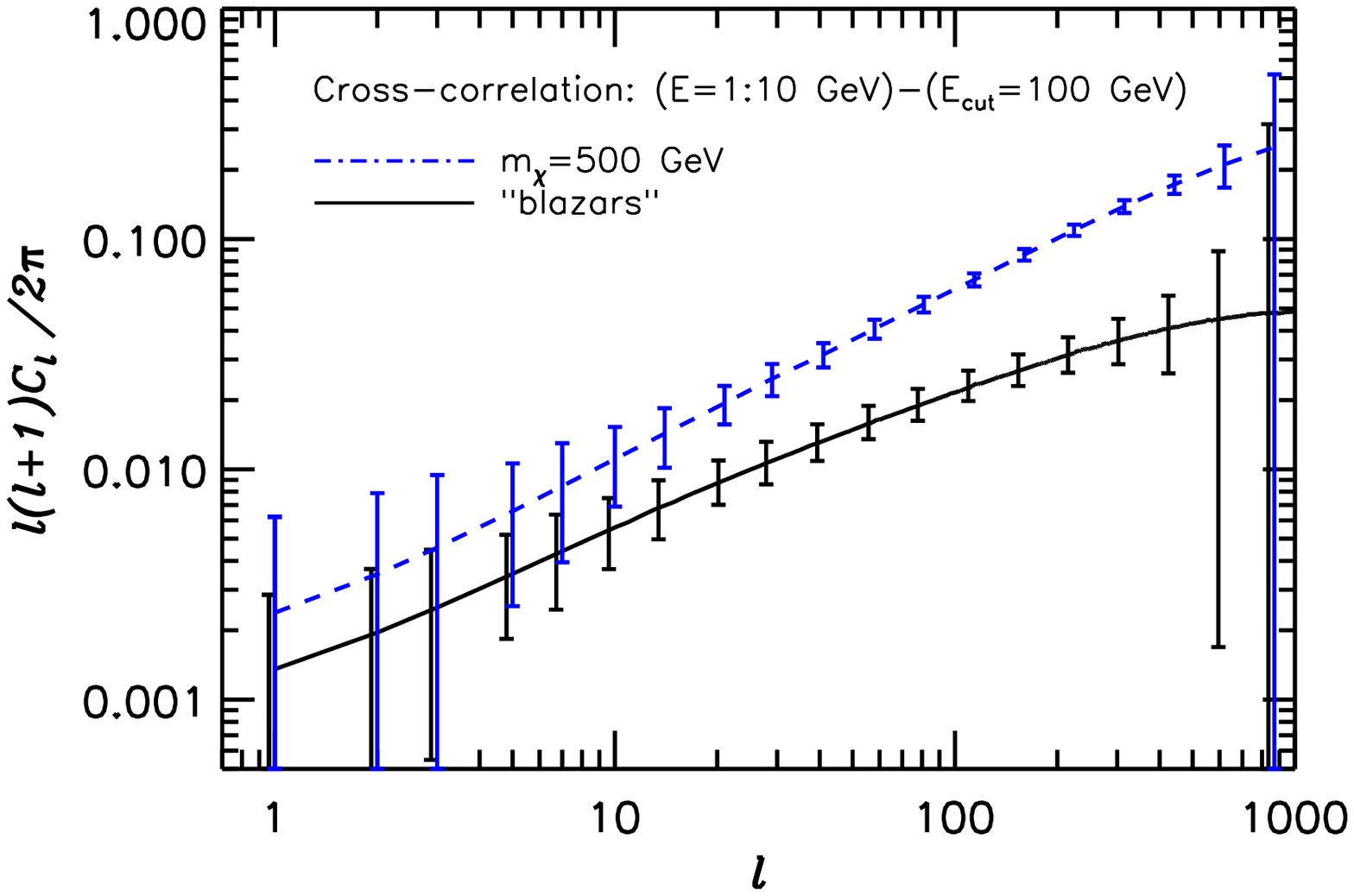}
\end{tabular}
\end{center}
\vspace{-2pc} \caption{Cross-correlation between the energy bands
\mbox{$E=1\!:\!10$ GeV} (\mbox{1 GeV$<\!E\!<$10 GeV}) and $E_{\rm
cut}=10,100$ GeV. The errors refer to a 4-year GLAST survey.
\label{fig:angularcrossspectra_eb} } \vspace{-1pc}
\end{figure}

The cross-correlation formalism introduced in the previous section
can be further employed in comparing the CGB anisotropies in
\emph{different energy bands}. In particular, at the low energy
band $E\alt 1-10$ GeV the DM contribution is expected to be
negligible. This energy range thus represents a natural, high
statistics template of the astrophysical gamma sky to compare with
for the higher, $E\agt 10$ GeV, DM relevant energy bands
considered above.

As an example  we plot in Fig.~\ref{fig:angularcrossspectra_eb}
the cross-correlation between the energy bands $E=1\!:\!10$ GeV
(i.e. \mbox{1 GeV$<\!E\!<$10 GeV}), where the DM contribution is
assumed to be negligible, and $E_{\rm cut}=10,100$ GeV as for
Figs.~\ref{fig:angularspectra2} and \ref{fig:angularspectra1} for
an astrophysical dominated CGB and for a DM dominated CGB for
various WIMP masses $m_\chi$. It can be seen that the diagnostic
power is similar to that of the auto-correlation of
Fig.~\ref{fig:angularspectra2}, understandable in the light of the
close similarity between the $E_{\rm cut}=10,100$ GeV CGB and the
template we are comparing with. The cross-correlation between
different energy bands thus  represents a further independent
observable sensitive to DM signatures. In particular, it acts
complementary to the auto-correlation spectrum, providing an
effective, high statistics, calibration of the astrophysical
background at low energy, thus allowing more easily to distinguish
the sought DM signal at higher energies.

%%%%%%%%%%%%%%%%%%%%%%%%%%%%%%%%%%%%%%%%%%%%
\section{Discussion}\label{discussion}
%%%%%%%%%%%%%%%%%%%%%%%%%%%%%%%%%%%%%%%%%%%%

\subsection{Mixed scenario}

We have seen that a particularly clear signature of DM
annihilation in the CGB is present in our ``quadratic scenario''.
However, a certain contribution from sub-galactic clumps and thus
a mixing of the linear and quadratic scenarios is anyway expected,
although, as previously discussed, the relative contribution  is
still quite uncertain. A contribution to the DM signal from
sub-galactic clumps is particularly interesting due to the fact
that it is expected to enhance the overall DM annihilation signal
of roughly one order of magnitude increasing correspondingly the
chances of detection
\cite{Ullio:2002pj,Bergstrom:1998jj,Taylor:2002zd}. To give a hint
of how this contribution affects the previous conclusions we show
in Fig.~\ref{fig:angularspectracross} the auto-correlation
spectrum of DM for $E_{\rm cut}=100$ GeV and for a $m_\chi= 500$
GeV WIMP in the case in which 80\% of the DM signal correlate
linearly with matter and only 20\% of the DM contribution
correlate quadratically, the sum being at the level of the EGRET
flux. Given that DM in the linear scenario is almost degenerate
with a pure astrophisical emission, an equivalent interpretation
of Fig.~\ref{fig:angularspectracross} is that of a subdominant,
20\% level, quadratic DM contribution, and an overall signal
dominated by astrophysical emission. We see that in both cases the
prospects are quite interesting and the DM spectrum still differs
significantly from that of a background generated by astrophysical
sources only.

In the ``worst'' case, i.e. our ``linear scenario'', in which
sub-galactic clumps dominate the annihilation signal, the
anisotropies are degenerate with the astrophysical signal and the
signature in the CGB disappear. In second approximation some
difference is still expected due to the presence of a bias in the
relative distributions of DM and astrophysical sources, although
the signature become quite model dependent (see \cite{Ando:2006cr}
for a more detailed discussion). However, in this case, unless our
galaxy is unrepresentative of an average galactic halo, the best
chances to detect the DM gamma signal, clearly, would come from
the Milky Way halo itself for which other kinds of anisotropy
signatures, due basically to the peculiar profile of the galactic
halo, are expected (see for example
\cite{Pieri:2007ir,Hooper:2007be} for more details). In this
respect, it is interesting to notice this sort of complementarity
between DM signatures in the extragalactic cosmological signal and
the local galactic signal.

\subsection{CGB  normalization}

\begin{figure}[t]
\vspace{-1pc}
\begin{center}
\includegraphics[width=0.50\textwidth]{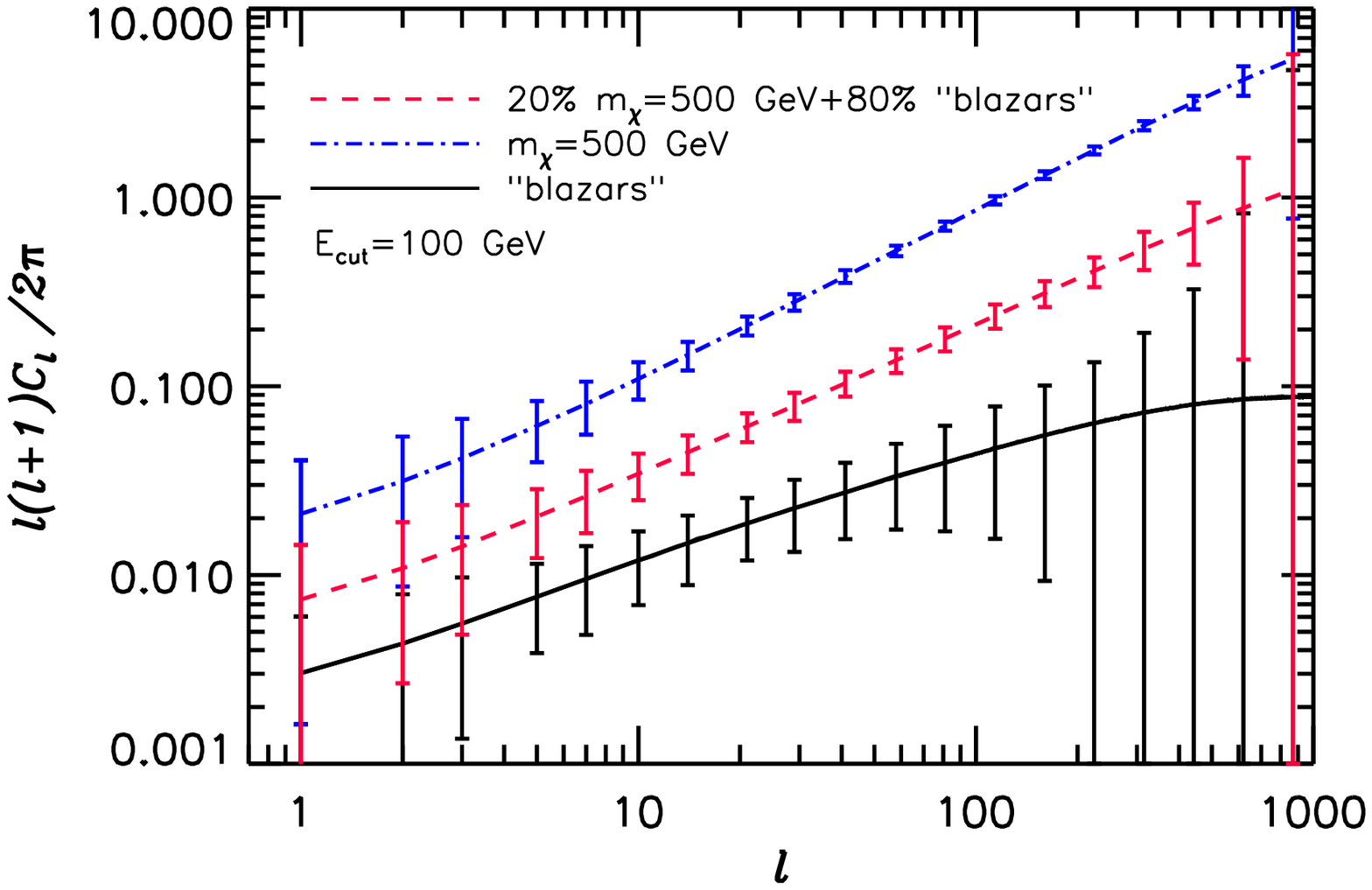}
\vspace{-1.5pc} \caption{Angular spectra for $E_{\rm cut}=100$
GeV. Shown are the cases of  CGB dominated by ``blazars'',  CGB
dominated by a $m_\chi= 500$ GeV WIMP and  CGB contributed by a
$m_\chi= 500$ GeV WIMP for 20\% and by ``blazars'' for 80\%. This
latter case is degenerate with a CGB contributed entirely by DM
with a 20\% emission tracing the matter quadratically  and an 80\%
emission tracing the matter linearly. The errors refer to the
statistics expected from a 4-year GLAST survey.
\label{fig:angularspectracross} } \vspace{0pc}
\begin{tabular}{c}
\includegraphics[width=0.50\textwidth]{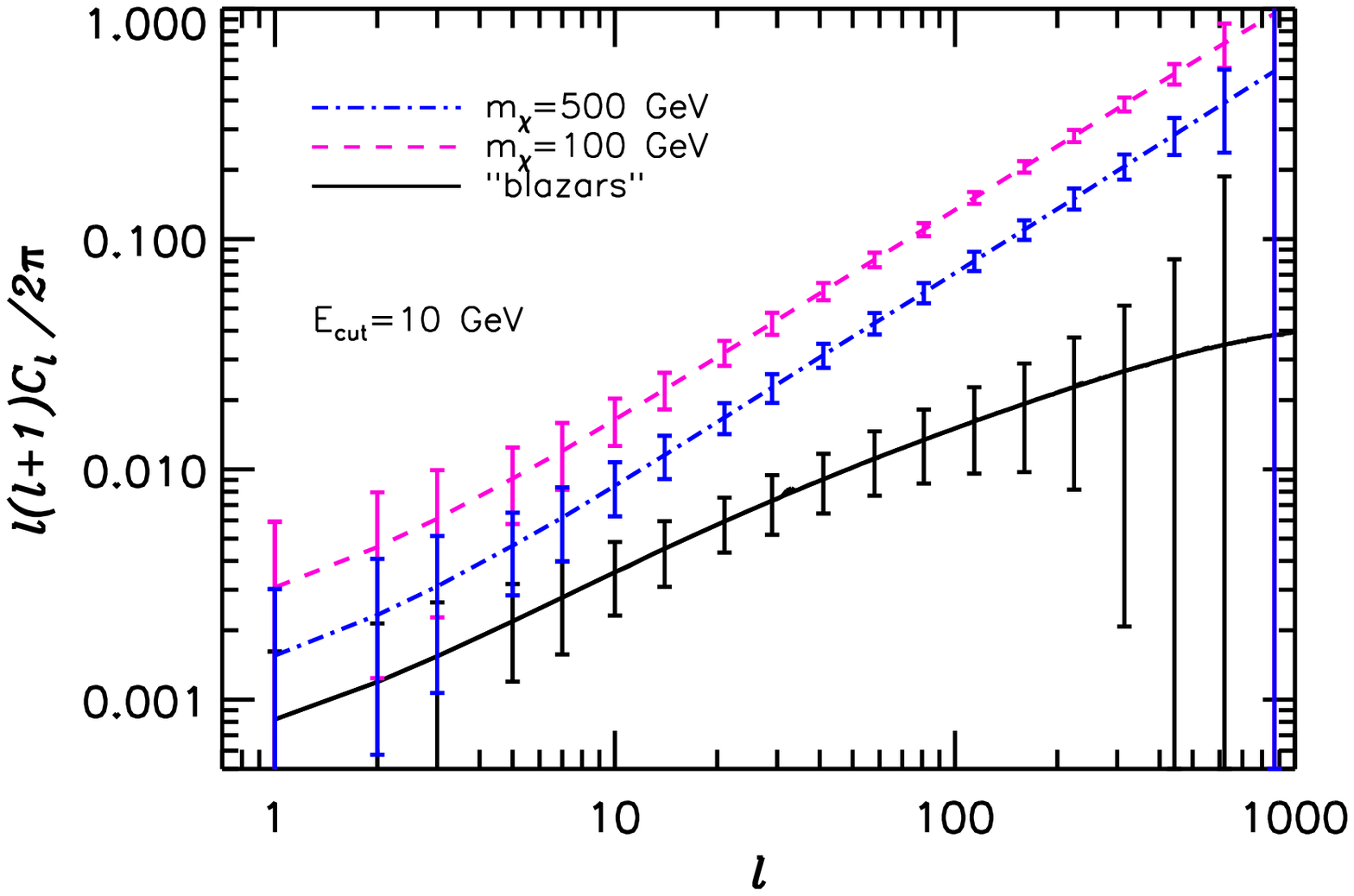} \vspace{-1.0pc} \\
\includegraphics[width=0.50\textwidth]{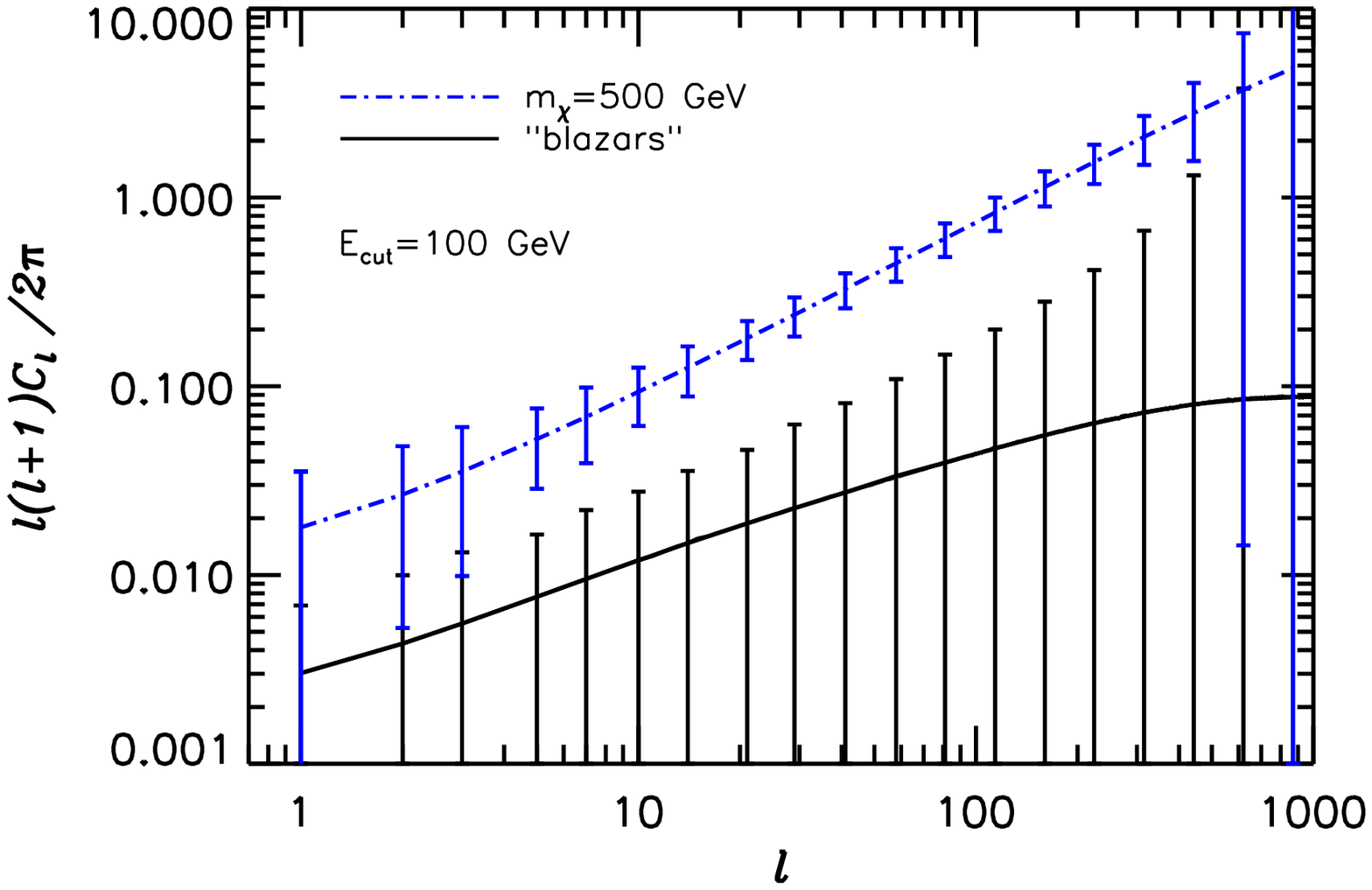}
\end{tabular}
\end{center}
\vspace{-1.5pc} \caption{As in Fig.~\ref{fig:angularspectra2}, but
for a DM and astrophysical signal 5 times lower (i.e~20\% of the
present EGRET value). \label{fig:angularspectralow} }
\vspace{-1pc}
\end{figure}

Part of the population of sources contributing to the CGB will
likely be resolved by GLAST consequently lowering the level of
unresolved emission and thus the intensity of the CGB. This,
indeed, will turn out as an advantage given that only
astrophysical sources are resolved and thus the signal to noise
ratio for DM is \emph{enhanced}. An estimate in the framework of
the blazar model of the CGB of \cite{Stecker:1996ma} suggests that
GLAST could lower the CGB by a factor of 2 \cite{Ullio:2002pj}. As
an extreme assumption we plot in Fig.~\ref{fig:angularspectralow}
the error bars in the case in which the CGB (and thus the
statistics) is reduced by a factor of 5 (i.e.~to 20\% of the
present value), assuming the pure quadratic DM scenario. We see
that even in this case the statistics are good enough to separate
the two angular spectra. Notice that the result is quite
conservative given that in the figure both the CGB and the DM
signal are reduced by a factor of 5.

Although not shown, a very similar result applies for the case of
a cross-correlation between different CGB energy bands that thus
equally maintain its diagnostic power in a low statistics regime.
The sensitivity of the galaxies-CGB cross-correlation is instead
sensibly reduced both in the low and the high energy ranges.
Finally, if we consider the mixed scenario in the framework of
this low statistics CGB then the prospects of DM detection became
quite low. A 20\% DM quadratic contribution in this case would
correspond to a 4\% contribution with respect to the present EGRET
intensity, making the anisotropy transition signature quite
challenging to detect.

\subsection{Foreground removal}

Finally, we comment on the role of the galactic foreground on the
results. The foreground subtraction remains a delicate issue, as
can be appreciated by the reanalysis of the EGRET data performed
in \cite{Strong:2004ry}, based on a revised model for the galactic
propagation of cosmic rays that resulted in an appreciably
different spectral behavior compared to Eq.~(\ref{spectrum98}) and
in a slight change in the overall normalization. The foreground
subtraction is also expected to alter and enlarge in a non-trivial
way the estimate in Eqs.~(\ref{gausserrors1}-\ref{gausserrors2})
of the error bars of the angular spectrum. A detailed estimate of
this effect is beyond the scope of the present work. Possibly,
however, the effect of galactic contamination can  be kept under
control enlarging the galactic cut to higher galactic latitudes
$b\agt20^\circ$ where the galactic emission is expected to rapidly
decrease \cite{Sreekumar:1997un,Strong:2004ry}, although at the
price of reducing correspondingly the available statistics.
Residual foregrounds are anyway expected even at the highest
latitudes, at a level depending on the foreground model used,
making non-obvious also this simple first order analysis. An
accurate analysis, further, should eventually rely on a full
simulation of the data analysis pipeline.

Further, if, as considered above, clumpiness in the Milky Way halo
becomes relevant, then, in principle, the resulting DM
annihilation signal has to be considered as a further galactic
foreground. CGB extraction, in this case, would become more
challenging due to the need to include in consistent way  DM
annihilation both in the galactic and the extra-galactic signal
(see, indeed, ref.~\cite{deBoer:2007zc} where an iterative
procedure is applied both to the galactic foreground and the
extra-galactic background for the claim of DM detection in the
EGRET data).

%%%%%%%%%%%%%%%%%%%%%%%%%%%%%%%%%%%%%%%%%%%%
\section{Summary and Conclusions}\label{final}
%%%%%%%%%%%%%%%%%%%%%%%%%%%%%%%%%%%%%%%%%%%%

In the present work we have studied the kind of signatures that DM
annihilation is expected to imprint in the anisotropies of the
CGB, complementary to the signatures in the energy spectrum. We
have addressed the main physical ingredients contributing to the
DM signature and discussed the robustness of the signature with
respect to various possible scenarios. We can summarize our
findings as follows:
\begin{itemize}
     \item{The DM annihilation signal traces in general
     the matter distribution quadratically due to its $\rho_\chi^2$
     dependence. However, an effective linear correlation can arise if the
     signal is significantly enhanced by the presence of cuspy
     haloes or sub-galactic clumps. We have defined the two extreme ``linear'' and ``quadratic''
     scenarios. The first corresponds to the case in which the cosmological DM annihilation
     signal is dominated by galactic or sub-galactic structures while in the second the signal
     is dominated by emission on scales larger than that of a galactic
     halo. We have chosen a phenomenological
     approach introducing a parameter $\xi$ that weights the two relative
     contributions exploring the DM signatures for different possible choices of $\xi$.}

     \item{The anisotropies are determined both by the intrinsic
     fluctuations in the source field and by the size of the emission horizon
     $z_{\mathcal{H}}$. For $E_{\rm cut}\agt 100$ GeV the horizon $z_{\mathcal{H}}$ is
     essentially fixed by photon absorption in the
     EBL.  The bulk of the gamma-rays is expected to originate inside
     $z_{\mathcal{H}}\approx 0.5$,
     independent of whether they have a DM or an astrophysical origin.
     For $E_{\rm cut}\agt 10$ GeV,  DM annihilation in the quadratic scenario
     has a redshift horizon $z_{\mathcal{H}}\approx
     3-4$. The horizon is still significantly limited by PP losses at this
     energy, otherwise exceeding $z_{\mathcal{H}}\approx
     10$. Blazars and DM in the linear scenario have degenerate horizons $z_{\mathcal{H}}\approx
     1$. }

     \item{ In the quadratic scenario the DM
     anisotropy signal is sensibly enhanced with respect to
     blazars for  $E_{\rm cut}=$10 GeV. Further, also the shapes
     of the angular spectra differ significantly
     \cite{Ando:2005xg,Ando:2006cr}.
     The signature remain standing also  for $E_{\rm cut}=$ 100 GeV despite the
     decreased statistics and become particularly strong,
     being independent of uncertainties related to the
     blazar-matter bias or to the evolution of blazars. This
     scenario can easily be detected by GLAST and would constitute
     a strong signature of DM annihilation. The DM linear scenario, instead, exhibit the same level
     of fluctuations of blazars and the two thus have almost degenerate anisotropy features.}

    \item{The above signature in the angular spectrum remains quite
    robust as long as the the quadratic DM
    signal is at least at the 10-20\% level with respect to the
    linear DM or the blazar component. A further uncertainty to
    take into account is the normalization of the CGB (and thus
    the available statistics) that is likely to be reduced if part of the sources
    contributing to the CGB will be resolved by GLAST. If the normalization is reduced
    by an extreme factor of 5, (20\% of the present EGRET value),
    the pure quadratic DM scenario exhibits still a relevant
    anisotropy transition signature. If the quadratic DM
    contributes for a 20\% (thus, 4\% of the present EGRET value) then
    the detection of the signature becomes quite challenging.}

    \item{The cross-correlation between the CGB and  a survey of
    galaxies and  the cross-correlation between different energy bands of the CGB
    provide further independent and sensitive
    observables that can be employed in combination with the CGB
    auto-correlation. A joint analysis of all the anisotropy
    observables considerably improves the sensitivity to the DM signal
    and, more in general, the power of the statistical
    analysis. In principle,
    the exact contribution from DM annihilation in sub-galactic
    clumps and cuspy haloes  can be
    treated as free parameters (instead of relying on a model)
    and inferred from the analysis.}

\end{itemize}

The above conclusions hold exactly if a perfect cleaning of the
galactic foregrounds and a lossless extraction of the CGB signal
is possible. The analysis of foregrounds will be likely the main
challenge in the study of the CGB. Clearly, given the above shown
potential of CGB anisotropies in looking for DM signatures, it
would be worth to perform further detailed studies on the issue.
The launch of the GLAST satellite is expected by the middle of
2008, while the satellite AGILE \cite{AGILE} launched in April
2007 is currently already taking data. The improvement in
statistics compared to EGRET will allow for new, powerful tools to
search for exotic contributions to the gamma-ray signal.  The
anisotropy analysis of the CGB in particular, if foregrounds
contaminations can be efficiently kept under control, promises to
provide a clear signature of DM annihilation or, in the case of a
negative answer, to obtain new constraints on the DM properties,
complementary to a pure energy spectrum analysis.

\vspace{1pc}
%%%%%%%%%%%%%%%%%%%%%%%%%%%%%%%%%%%%%%%%%%%%
\section*{Acknowledgments}
%%%%%%%%%%%%%%%%%%%%%%%%%%%%%%%%%%%%%%%%%%%%

The authors wish to thank P.~D.~Serpico for fruitful discussions and
for valuable comments on the manuscript. H.~Tu is kindly
acknowledged for providing us the Halo-model power spectrum. We
thank the Danish Centre of Scientific Computing (DCSC) for granting
the computer resources used. TH acknowledges partial financial
support from the Spanish Research Ministry (MEC), under the contract
FPA2006-05807. GM acknowledges supports by Generalitat Valenciana
(ref.\ AINV/2007/080 CSIC) and by PRIN 2006 ``Fisica
Astroparticellare: neutrini ed universo primordiale'' by the Italian
MIUR.

%%%%%%%%%%%%%%%%%%%%%%%%%%%%%%%%%%%%%%%%%%%%
%%%%   Appendix
%%%%%%%%%%%%%%%%%%%%%%%%%%%%%%%%%%%%%%%%%%%%
%\appendix
%%%%%%%%%%%%%%%%%%%%%%%%%%%%%%%%%%%%%%%%%%%%
%%%%%%%%%%%%%%%%%%%%%%%%%%%%%%%%%%%%%%%%%%%%
%%%%   References
%%%%%%%%%%%%%%%%%%%%%%%%%%%%%%%%%%%%%%%%%%%%

%%%%%%%%%%%%%%%%%%%%%%%%%%%%%%%%%%%%%%%%%%%%
\end{document}